\documentclass[11pt]{article}

\usepackage{amsmath,amssymb}
\usepackage{graphicx,color}
\usepackage{mathrsfs}

\def\tr{{\,\mathrm{tr}\,}}

\numberwithin{equation}{section}
\begin{document}

\quad 
\vspace{-2.0cm}

\begin{flushright}
\parbox{3cm}
{
{\bf July. 2010}\hfill \\
YITP-10-57 \hfill \\
 }
\end{flushright}

\vspace*{2.5cm}

\begin{center}
\huge\bf 
{Surface Operator, Bubbling Calabi-Yau \\and AGT Relation
}\\
\end{center}
\vspace*{1.5cm}
\centerline{\Large
{Masato Taki}
}
\vspace*{-0.2cm}
\begin{center}
\large{\textit{Yukawa Institute for Theoretical Physics, Kyoto University, Kyoto 606-8502, Japan}}
\vspace*{-0.2cm}
\begin{center}
\texttt{taki@yukawa.kyoto-u.ac.jp}\\
\end{center}

\end{center}

\vspace*{0.7cm}

Surface operators in $\mathcal{N}=2$ four-dimensional gauge theories are interesting half-BPS objects.
These operators inherit 
the connection of gauge theory with the Liouville conformal field theory, 
which was discovered by Alday, Gaiotto and Tachikawa.
Moreover it has been proposed that 
toric branes in the A-model topological strings 
lead to surface operators via the geometric engineering.
We analyze the surface operators
by making good use of topological string theory.
Starting from this point of view, 
we propose that the wave-function behavior of the topological open string amplitudes
geometrically engineers the surface operator partition functions 
and the Gaiotto curves of corresponding gauge theories.
We then study a peculiar feature
that the surface operator corresponds to the insertion of the degenerate fields
in the conformal field theory side.
We show that this aspect can be realized as the geometric transition in topological string theory,
and the insertion of a surface operator leads to the bubbling of the toric Calabi-Yau geometry.

\vspace*{0.5cm}

\vfill

\thispagestyle{empty}
\setcounter{page}{0}

\newpage

\section{Introduction}
The $\mathcal{N}=2$ four-dimensional gauge theories 
have provided many insights into the study of dynamics of gauge theory and string theory.
In particular Seiberg and Witten \cite{Seiberg:1994rs,Seiberg:1994aj} found that
the low-energy dynamics of these theories is governed by the geometry of Riemann surfaces.
Moreover Nekrasov \cite{Nekrasov:2002qd} gave a microscopic derivation of the Seiberg-Witten geometry.

Last year, a mysterious relation
which relates $\mathcal{N}=2$ gauge theories in four dimensions
to conformal field theories in two dimension was discovered in \cite{Alday:2009aq}.
This relation originated in the M5-brane construction of $\mathcal{N}=2$ gauge theories which was
developed in \cite{Gaiotto:2009we}.
In this construction, a four-dimensional $\mathcal{N}=2$ superconformal gauge theory arises from compactification of
M5-branes on a Riemann surface with punctures.
Since a $\mathcal{N}=2$ superconformal gauge theory of such a type is associated with a Riemann surface,
it is very natural that 
there exists a relation between the Seiberg-Witten-Nekrasov solution of a $\mathcal{N}=2$ superconformal gauge theory and the conformal field theory on the corresponding surface.
Inspired by the development, Alday, Gaiotto and Tachikawa found the specific realization of the idea, 
which is called the AGT relation \cite{Alday:2009aq}:
the Nekrasov partition function of a $\mathcal{N}=2$ superconformal gauge theory 
is equal to the Liouville conformal block for the Riemann surface.

The part of the extended objects, such as 'tHooft-Wilson line operators and surface operators, 
in the dual Liouville theory was explained in \cite{Drukker:2009tz,Alday:2009fs,Drukker:2009id}:
the $\mathcal{N}=2$ gauge theory in the presence of an elementary surface operator
corresponds to the Liouville conformal block with an added insertion of a
degenerate primary field $\Phi_{2,1}(z)$.
The expectation values of  'tHooft-Wilson line operators are then computed 
by using the monodromy operation for the conformal blocks.
In this way the AGT relation can be extended to gauge theories in the presence of 
extended observables.

In this paper we study the stringy realization of the AGT relation in the presence of elementary surface operators.
By using M2-brane construction of surface operators \cite{Alday:2009fs},
we can obtain surface operators from toric A-branes of topological string theory 
via the Ooguri-Vafa duality \cite{Ooguri:1999bv}.
This is an extended version of the well-known geometric engineering 
of $\mathcal{N}=2$ gauge theories in four-dimensions \cite{Katz:1996fh}.
The open topological string construction enables us to grasp some important properties of the AGT relation.

In \cite{Alday:2009aq} it was found that the degenerate conformal block in the semi-classical limit 
gives the Seiberg-Witten curve in the Gaiotto form.
We interpret this property string-theoretically 
by using the wave-function behavior of the open topological strings \cite{Aganagic:2003qj}
and mirror symmetry.
Since this approach gives the differential equation for the expectation value of a surface operator,
we can compute it very efficiently.
Then we find that a Gaiotto curve relates to mirror Calabi-Yau corresponding to a $\mathcal{N}=2$ gauge theory
via four-dimensional limit or decoupling one of gravity.

Recall the extended AGT relation \cite{Alday:2009fs} that
the insertion of the degenerate field leads to 
the Nekrasov partition function in the presence of a surface operator,
which we call the ramified instanton partition function.
The degenerate conformal block is equal to a certain Nekrasov partition function
for the specialized masses via the AGT relation.
These facts imply that the Nekrasov partition function of a gauge theory 
gives the ramified instanton partition function of another gauge theory
by degenerating physical mass parameters.
We show that this is precisely the geometric transition of the A-model topological strings
under which branes change to the bubbling of a background Calabi-Yau geometry.

The organization of this paper is as follows:
In section 2 we study surface operators in the context of topological string theory.
The role of the wave-function behavior in the AGT relation becomes clear.
In section 3 we discuss the extended AGT conjecture for the insertion of surface operators
from the viewpoint of topological string theory.
We explain why the degenerate field insertion leads to the inclusion of a surface operator in the gauge theory side.
The geometric transition gives a partial explanation to the question.
In Appendix A we present a brief review on the Young diagrams and the Schur functions.
We review the five-dimensional Nekrasov partition functions and the refined vertex in Appendix B.

  While preparing this paper, \cite{Dimofte:2010tz} 
  which has some overlap with this paper appeared.

\section{Surface Operators and Topological Strings}
In this section we study the simple surface operators of
$\mathcal{N}=2$ four dimensional gauge theories
via the geometric engineering.
We illustrate the relation between toric branes and surface operators
with a simple example, which is called the $\mathcal{T}_{3,0}$ theory.
This theory was studied extensively in \cite{Wyllard:2009hg,Schiappa:2009cc,Kozcaz:2010af}.
We show that the Gaiotto curve for the gauge theory
comes of the wave function behavior of the corresponding open topological string amplitude.

\subsection{Toric brane and surface operator for $\mathcal{T}_{3,0}$ theory}

\begin{figure}[htbp]
 \begin{center}
  \includegraphics[width=100mm,clip]{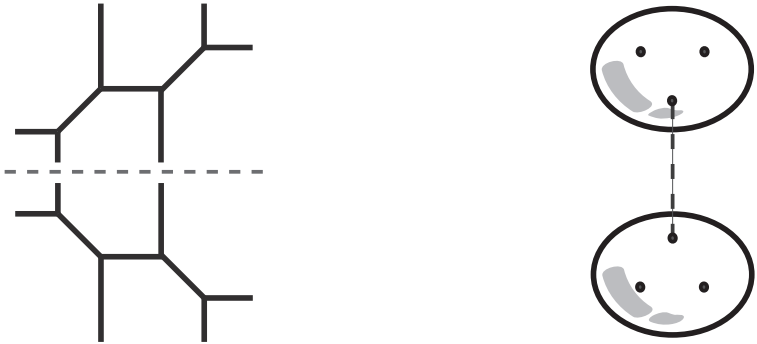}
 \end{center}
 \caption{The toric geometry which engineers the $SU(2)$ gauge theory with four flavors (left)
 and the corresponding Gaiotto curve (right).
 This geometry is constructed by gluing two geometries that engineers the four free hypermultiplets theory
 respectively.
 }
 \label{fig:4flav}
\end{figure}

$\mathcal{T}_{3,0}(A_1)$ theory, which is just the theory of four free hypermultiplets,
are associated with the Riemann sphere with three punctures via the AGT relation.
In this section, we study how surface operators arise from string theory setup.
The key to stringy realization is the geometric engineering of $\mathcal{N}=2$ gauge theories.
In \cite{Katz:1996fh} it was shown that
Seiberg-Witten solutions are engineered from the genus-zero topological string theory on
the corresponding Calabi-Yau geometries.
The higher genus extension of the engineering was studied extensively in
\cite{Iqbal:2002we,Iqbal:2003ix,Iqbal:2003zz,Eguchi:2003sj,Hollowood:2003cv,Eguchi:2003it},
and they found that 
all-genus topological string partition function
leads to the Nekrasov partition functions through the topological vertex formalism \cite{Aganagic:2003db}.
The toric geometry which engineers the $SU(2)$ gauge theory with four flavors
is shown in the left side of  Fig.\ref{fig:4flav}.

Let us compactify M5-branes on a Riemann surface 
in order to construct $\mathcal{N}=2$ gauge theory \cite{Witten:1997sc,Gaiotto:2009we}.
The $SU(2)$ gauge theory with four flavors $\mathcal{T}_{4,0}(A_1)$
comes of the Riemann surface with four punctures.
We turn off the gauge coupling to get free theory:
this is done by tearing the surface into spheres with three punctures.
Since the Riemann sphere with four punctures 
degenerates into two Riemann surfaces with three punctures,
we obtain the toric geometry associated with this $\mathcal{T}_{3,0}(A_1)$ theory
by decomposing the toric geometry of $\mathcal{T}_{4,0}(A_1)$ as Fig.\ref{fig:4flav}.
We thus employ the half-geometry of Fig.\ref{fig:4flav}
as the toric geometry for the $\mathcal{T}_{3,0}$ theory.

\begin{figure}[htbp]
 \begin{center}
  \includegraphics[width=70mm,clip]{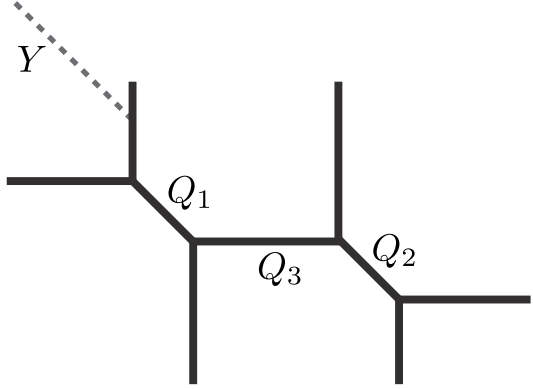}
 \end{center}
 \caption{The open A-model topological strings with
a toric A-brane on the leg. The background geometry is the toric Calabi-Yau three-fold which engineers
the $\mathcal{T}_{3,0}$ theory.
The dotted line here denotes the non-compact Lagrangian branes. 
The Young diagram $Y$ labels the representations of the Wilson loop for the boundary of the world-sheet.}
 \label{fig:T30open}
\end{figure}

Let us consider the open A-model topological strings on this geometry in the presence of
a toric A-brane on a non-compact Lagrangian submanifold.
Let $U$ be the holonomy matrix for the non-dynamical gauge field on the A-brane.
The open A-model amplitude for Fig.\ref{fig:T30open} is then given by
\begin{align}
&Z^{\textrm{open}}=
\sum_{Y} \textrm{Tr}_Y U \cdot Z_Y(Q_1,Q_2,Q_3\,; q),
\end{align}
and it is easy to compute $Z_Y(Q_1,Q_2,Q_f\,; q)$ by using the topological vertex \cite{Aganagic:2003db}.
We will compute the refined version of the amplitude in the next section.
Since we insert a single A-brane, the holonomy is one by one matrix $U=\textrm{diag}(z)$.
The trace for the representation $R$ 
is then non-zero only for the Young diagrams with one column $Y=[1^n]$,
and we obtain the following simple expression:
\begin{align}
&Z^{\textrm{open}}(z)=
\sum_{n=0}^\infty z^n \cdot Z^{(n)}(Q_1,Q_2,Q_3\,; q),\\
& Z^{(n)}=\prod_{i=1}^n 
\frac{(1-q^{i-1}Q_1)(1-q^{i-1}Q_1Q_2Q_3)}
{(1-q^i)(1-q^{i-1}Q_1Q_3)}.
\end{align}
We study the gauge theory interpretation of this open string amplitude.

\subsection{Gaiotto curves via wave function behavior}

Let us introduce the following parametrization:
\begin{align}
z=e^{-u},
\qquad q=e^{-g_s},
\end{align}
where $g_s$ is the topological string coupling constant.
Since $q^{-\partial_u}$ is the shift operator $q^{-\partial_u}e^{-u}=qe^{-u}$, 
it acts on the partition function as follows:
\begin{align}
(1-q^{-\partial_u})\cdot e^{-nu} Z^{(n)}
=(1-q^{n})\cdot e^{-nu} Z^{(n)}
=
\frac{(1-q^{n-1}Q_1)(1-q^{n-1}Q_1Q_2Q_3)}
{(1-q^{n-1}Q_1Q_3)}
e^{-u}e^{-(n-1)u} Z^{(n-1)}.\nonumber
\end{align}
Therefore, the operator $1-q^{-\partial_u}$ shifts the length of a column $[1^n]$.
This relation gives the following differential equation for the open string partition function
\begin{align}
\label{waveeq}
\left((1-Q_1Q_3q^{-\partial_u-1})(1-q^{-\partial_u})
-z{(1-Q_1q^{-\partial_u})(1-Q_1Q_2Q_3q^{-\partial_u})}
\right)Z^{\textrm{open}}(z)=0.
\end{align}
This characteristic of the open topological string amplitudes
is called the wave function behavior \cite{Aganagic:2003qj,KashaniPoor:2006nc,KashaniPoor:2008xg}.
This feature relates to the mirror B-model of the system:
the mirror of A-brane is the fermion $\psi(z)$
in the two-dimensional Kodaira-Spencer theory.
Then quantization of such a mirror system on a Riemann surface
leads to the wave equation $H\cdot \langle \psi \rangle=0$. 
The Hamiltonian, or the Schr\"oringer difference operator $H$, 
is associated with the local mirror Calabi-Yau geometry $xy+H(u,v)=0$.
By applying the wave equation for the following genus expansion of the wave function
\begin{align}
Z^{\textrm{open}}(z)=
\exp\left[\,-\left(
\frac{1}{g_s}W(z)+W_{(1)}(z)+g_sW_{(2)}+\cdots\right)
\right],
\end{align}
where each term $W_{(g)}$ is determined recursively.
We study the recursion relation from the viewpoint of the geometric engineering of the gauge theories.

\subsection*{Field theory limit }
Let $\hbar$ be the self-dual $\Omega$-background in four dimensions
$\epsilon_1=-\epsilon_2=\hbar$.
Then the gauge theory parameters are introduced by
\begin{align}
q=e^{-R\hbar },\qquad Q_a=e^{-Rm_a},
\end{align}
where $R$ and $m_a$ are the fifth-dimensional radius and masses respectively.
We take the four-dimensional limit of the open string amplitude as follows:
\begin{align}
\mathcal{W}_{\,(n)}(z)=\lim_{R\to 0}\,R^{n-1}\,W_{(n)}(z),
\end{align}
where $W_{(0)}=W$.
Then the four-dimensional partition function, which corresponds to the Nekrasov partition function
in the presence of a surface operator, takes the following form:
\begin{align}
\label{T304D}
\Psi (z)=
\lim_{R\to 0}\,Z^{\textrm{open}}(z)&=
\exp\left[\,-\left(
\frac{1}{\hbar}\mathcal{W}(z)+\mathcal{W}_{(1)}(z)+\hbar\mathcal{W}_{(2)}+\cdots\right)
\right].\nonumber\\
&=\sum_{n=0}^{\infty}
z^n
\prod_{j=1}^n 
\frac{(j\hbar-\hbar+m_1)(j\hbar-\hbar+m_1+m_2+m_3)}
{j\hbar(j\hbar-\hbar+m_1+m_3)}.
\end{align}
Notice that this is a fundamental solution of the hypergeometric differential equation near $z=0$
\begin{align}
{}_2F_1\left(\frac{m_1}{\hbar},\frac{m_1+m_2+m_3}{\hbar}; \frac{m_1+m_3}{\hbar};z\right).
\end{align}
This coincides with the sphere conformal block with the insertion of two primary fields
and a degenerate field.

We want to extract the information of the Seiberg-Witten curve, or Gaiotto curve,
from the partition function.
The wave function behavior then plays a key role.
Let us expand the differential equation in $R$
\begin{align}
&\big((m_1+m_3 -\hbar\partial_u-\hbar+\cdots)(-\hbar\partial_u+\cdots)\nonumber\\
&\qquad\quad -z{(m_1-\hbar\partial_u+\cdots)
(m_1+m_2+m_3-\hbar\partial_u+\cdots)}
\big)Z^{\textrm{open}}(z)=0,
\end{align}
where the dots $\cdots$ represent terms of order $\mathcal{O}(R)$.
In the field theory limit, the wave equation
(\ref{waveeq}) takes the following form
\begin{align}
\label{HGEfotT30}
\big[
(1-z)\hbar^2{\partial_u}^2
-&\left((m_1+m_3-\hbar)-z(2m_1+m_2+m_3)
\right)\hbar\partial_u \nonumber\\
&\qquad\qquad\qquad\qquad
-zm_1
(m_1+m_2+m_3)
\big]\Psi(z)=0.
\end{align}
This four-dimensional open string wave function is exactly the hypergeometric differential equation
whose solution is (\ref{T304D}).
The lesson here is that the wave function behavior of the open topological strings
gives the null state condition on the degenerate conformal blocks 
which are the ramified instanton partition functions via the AGT relation.
This construction, or the geometric engineering, 
of the the ramified Nekrasov partition functions implies 
that the Gaiotto curve can be derived from the mirror Riemann surface
via geometric engineering.

The wave function (\ref{HGEfotT30}) in the leading order $\hbar^0$ constraints the superpotential $\mathcal{W}$,
which is four-dimensional limit of the disk amplitude, as follows
\begin{align}
(1-z){\partial_u\mathcal{W}(z)}^2
+\left((m_1+m_3)-z(2m_1+m_2+m_3)
\right)
\partial_u\mathcal{W}(z) 
-zm_1
(m_1+m_2+m_3)
=0.
\end{align}
It is easy to solve this ``loop equation" as follows
\begin{align}
\mathcal{W}(z)
=\alpha_2\log z
+\alpha_3\log (1-z)
\pm\int^{z}\frac{\sqrt{\alpha_1^2({z^\prime}^2-z^\prime)
+\alpha_2^2(1-z^\prime)  
+\alpha_3^2z^\prime
}}
{z^\prime(1-z^\prime)}dz^\prime
,
\end{align}
where we introduce the CFT parameters
\begin{align}
m_1=-\alpha_1+\alpha_2-\alpha_3,\quad\,\,
m_2=\alpha_1-\alpha_2-\alpha_3,\quad\,\,
m_3=\alpha_1+\alpha_2+\alpha_3.
\end{align}
The logarithmic terms of the solution are just the overall factor $z^{\alpha}(1-z)^{\alpha^\prime}$.
The Gaiotto curve is then given by $\partial_u\mathcal{W}_{(0)}=\phi_2(z)$ as follows:
\begin{align}
\phi_2(z)
=\frac{\sqrt{\alpha_1^2(z^2-z)
+\alpha_2^2(1-z)  
+\alpha_3^2z}}
{z(1-z)}.
\end{align}
This is the well-known Gaiotto curve for the sphere with three punctures, 
or the $\mathcal{N}=2$ theory of four free hypermultiplets $\mathcal{T}_{3,0}$ 
\cite{Wyllard:2009hg,Schiappa:2009cc}.

\subsection*{Generic $\Omega$-background and Nekrasov-Shatashivili limit}
We can also compute the wave function 
for the generic $\Omega$-background $\epsilon_1\neq -\epsilon_2$.
We call the corresponding extension of topological string theory
``the refinement" \cite{Iqbal:2007ii}.
The refined topological vertex computation in the next section leads to 
the refined partition function
\begin{align}
&Z^{\textrm{open}}(z)=
\sum_{n=0}^\infty z^n 
\prod_{i=1}^n 
\frac{(1-q^{i-1}Q_1)(1-q^{i-1}Q_1Q_2Q_3)}
{(1-q^i)(1-q^{i}t^{-1}Q_1Q_3)}.
\end{align}
The difference equation follows from this expression of the refined partition function:
\begin{align}
\left((1-Q_1Q_3q^{-\partial_u}t^{-1})(1-q^{-\partial_u})
-z{(1-Q_1q^{-\partial_u})(1-Q_1Q_2Q_3q^{-\partial_u})}
\right)Z^{\textrm{open}}(z)=0.
\end{align}
This difference equation is the five-dimensional lift of the differential equation
derived from the null condition of the degenerate field \cite{Schiappa:2009cc,Kozcaz:2010af}.
It would be interesting to study such difference equations
by employing the five-dimensional version of the AGT relation
and the $q$-Virasoro conformal blocks \cite{Awata:2009ur}.

An important feature of the refined wave-equation is that
it depends almost exclusively on $q$ (or $t$) for the brane on an un-preferred toric leg\footnote[1]{In
this paper we study toric branes on an un-preferred leg of the refined topological vertex.
Branes placed on the preferred leg lead to the homological link invariants
through the geometric transition \cite{Gukov:2007tf}.}
\cite{Iqbal:2007ii,Gukov:2007tf}.
Recall that
$\epsilon_{1,2}$ are the rotational parameters of two distinct planes in $\mathbb{C}^2$,
and a sueface operator fills one plane.
It is therefore natural that the corresponding toric brane senses one-half of these parameters.
In the following, we study the wave-function in the Nekrasov-Shatashivili limit $t\to 1$ \cite{Nekrasov:2009rc}.
Now let us assume that the wave function in the limit
is almost equal to the unrefined one.
Then we can use the following mirror wave-equation for the toric brane \cite{Aganagic:2003qj}
in place of the differential equation in the Nekrasov-Shatashivili limit
\begin{align}
H(e^p,e^x)\cdot Z^{\textrm{open}}(z)=0,
\end{align}
where $[x,p]=g_s$.
Note that the polynomial $H(e^p,e^x)=0$ defines the mirror curve of the toric geometry.
We expect that this equation essentially leads to the Schr\"odinger equation for the corresponding integrable system.

Let  us take the $SU(2)$ pure super Yang-Mills theory for example.
The quadrangle toric web-diagram describes
the Calabi-Yau three-fold which engineers the gauge theory
 (see Fig.\ref{fig:mirrorA}).
Since we obtain the mirror geometry by fatting the toric diagram off,
the genus of the corresponding curve is one.
The local mirror symmetry \cite{Katz:1996fh} implies the defining equation
\begin{align}
z+\frac{1}{z}+y^2+c_1y+c_0=0,
\end{align}
where we fix the irrelevant coefficients by re-normalization.
The two K\"ahler moduli coordinates of B-model are given by the mirror map
\begin{align}
z_B=\frac{1}{{c_0}^2}=e^{-t_B}+\cdots,\quad z_F=\frac{1}{{c_1}^2}=e^{-t_F}+\cdots.
\end{align}
The quantum corrections of $z_B$, which is denoted by the dots,
are irrelevant in the field theory limit of interest,
since it is the large radius limit $c_0=1/\Lambda^2R^2$ and $R\to 0$.
We also fix
\begin{align}
u=\frac{1}{R^2}-\frac{\Lambda{c_1}^2}{4},
\end{align}
in order to recover the Seiberg-Witten curve:
\begin{align}
\Lambda^2\left(z+\frac{1}{z} \right)
+\left(y+\sqrt{\frac{1}{R^2}-u} \right)^2
+u=0.
\end{align}
Then the field theory limit $R\to 0$ implies
\begin{align}
\Lambda^2\left(z+\frac{1}{z} \right)
+p^2
+u=0.
\end{align}
where we take $y=-e^{-Rp}\sqrt{1/R^2-u}$.
In this limit, the quantization condition of the mirror curve becomes
$[x,p]=\epsilon_1$ where $z=e^x$.
This is precisely the Schr\"odinger equation for the sine-Gordon model
\begin{align}
\left(
{\epsilon_1}^2\frac{\partial}{\partial x}
+\Lambda^2\left(e^x+e^{-x}\right)
+u\right)\Psi=0.
\end{align}
In \cite{Mironov:2009uv,Mironov:2009dv,Popolitov:2010bz,Maruyoshi:2010iu},
the authors have proposed that the wave-functionof the sine-Gordon system
is equal to the ramified Nekrasov partition function in the Nekrasov-Shatashivili limit.
Our argument gives a naive derivation of the conjecture from the perspective of the geometric engineering.

Notice that we have assumed that the open string amplitude in the limit $t=1$
satisfies essentially the same equation as the self-dual case $q=t$.
It is not evident that such an observation is true for more complicated cases.
In fact the dual-CFT computation in \cite{Maruyoshi:2010iu}
shows that the differential equations for $SU(2)$ gauge theories
also depend on $\epsilon_2$.
This fact suggests that we have to modify our naive derivation.
Thus it is very important to improve our sketch of stringy derivation 
by formulating the refined version of the B-model and the corresponding integrable systems.

\subsection{Geometric engineering of surface operators }

So far, we have assumed that an open topological A-model amplitude gives
the corresponding Nekrasov partition function in the presence of a surface operator.
We then verify this expectation by combining the stringy construction of surface operators with string dualities.

Let us start with the M-theory construction of surface operators.
Recall that a gauge theory of \cite{Gaiotto:2009we} can be realized 
as the world-volume theory of M5-branes compactified on a
punctured Riemann surface.
A surface operators is then engineered by a semi-infinite transverse M2-brane.
Let $x,y\in\mathbb{C}$ be two complex combination of four coordinates $x^4, x^5,x^6$ and $x^7$.
Then the holomolorphic  curve which M5-branes wrap is described by $\Sigma:\,H(x,y)=0$.
We summarize the configuration as follows:
\begin{align}
\begin{array}{c|cccc|cccc|ccc}  &0 & 1 & 2 & 3 & 4 & 5 & 6 & 7 & 8 & 9 & 10  
\\ \textrm{M5} & {\bigcirc} & \bigcirc & \bigcirc & \bigcirc &  & \Sigma &  &  & \times & \times & \times
\\\textrm{M2} & \bigcirc & \bigcirc & \times & \times & \times & \times & \times & \times & \bigcirc & \times & \times\end{array}
\end{align}
Let us move on to Type IIA superstring setup.
The dimensional reduction along the M-circle $x^{10}$ leads to the type IIA string theory with
NS5-branes on $\Sigma$ and a transverse D2-brane.

\begin{figure}[htbp]
 \begin{center}
  \includegraphics[width=60mm,clip]{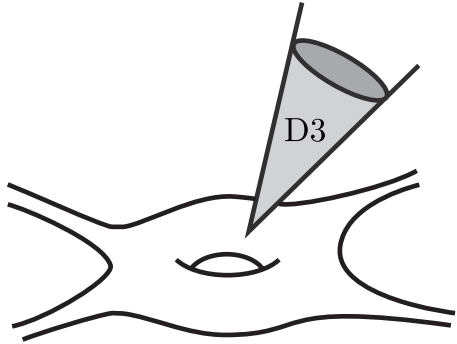}
 \end{center}
 \caption{The D3 brane wraps the non-compact two-cycle 
 whose circle fibration degenerates on the base Riemann surface $H(x,y)=0$.}
 \label{fig:mirror}
\end{figure}

Recall that the transversal T-duality replaces $N$ NS5-branes to a local $A_{N-1}$ singularity
\cite{Ruback:1986ag,Ooguri:1995wj}.
After T-dualizing, the circle fiber degenerates on the locus on which the NS5-branes were placed.
Thus, taking the T-duality along the circle transverse to $\Sigma$ \cite{Dijkgraaf:2007sw},
we obtain the type IIB string theory on the following Calabi-Yau geometry
\begin{align}
\label{mirrorCY}
uv+H(x,y)=0.
\end{align}
Then the D2 brane changes into a semi-infinite D3-brane by the T-duality.
This D3-brane wraps the following two-cycle parametrized a complex parameter $H(x_0,y_0)=0$
\begin{align}
\label{mirrorbrane}
u=0,\quad v=\textrm{arbitrary},\quad H(x_0,y_0)=0.
\end{align}
The cylinder fiber $uv=-H$ over the $xy$-plane degenerates on the curve $H(x,y)=0$ into
 two cones corresponding to $uv=0$,
and the D3-brane wraps one of them as illustrated in Fig.\ref{fig:mirror}.
The parameter $(x_0,y_0)$ is a modulus of the D-brane, 
which corresponds to a point on the base Riemann surface $\Sigma$.
This modulus parametrizes an elementary surface operator.

\begin{figure}[htbp]
 \begin{center}
  \includegraphics[width=40mm,clip]{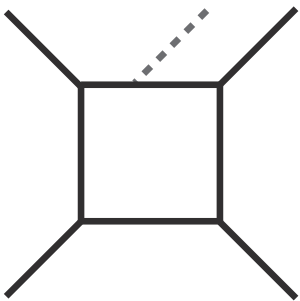}
 \end{center}
 \caption{The toric diagram of the Calabi-Yau mirror to $uv+H(x,y)=0$.
 The toric skeleton is just the thinned Riemann surface $H(x,y)=0$.}
 \label{fig:mirrorA}
\end{figure}

Finally we apply the mirror symmetry on this type IIB setup.
The mirror  of the geometry (\ref{mirrorCY}) is well-studied \cite{Hori:2000kt}:
it is a toric Calabi-Yau three-fold whose toric diagram is the skeleton 
which is obtained by thinning down the Riemann surface $\Sigma$.
The mirror brane of (\ref{mirrorbrane}) is also studied in \cite{Aganagic:2000gs,Aganagic:2001nx}:
it is the D4-brane which wraps the non-compact Lagrangian three-cycle
corresponding to the semi-infinite line on the toric diagram as Fig.\ref{fig:mirrorA}.
This D4-brane also fills the subspace of the space-time $\mathbb{R}^2\subset\mathbb{R}^4$.
This is precisely the setup of the Ooguri-Vafa duality \cite{Ooguri:1999bv}.
As was explained in \cite{Ooguri:1999bv}, 
we can compute the effective F-term in the presence of the D4-brane
by counting the open world-sheet instantons via the A-model open topological strings.
The open topological string amplitude for a toric Calabi-Yau such as Fig.\ref{fig:mirrorA}
is computed by using the topological vertex formalism.
Note that the topological vertex for the closed topological strings on the engineering toric Calabi-Yau
gives the corresponding Nekrasov partition function.
Therefore, by introducing the Lagrangian brane, we can extend this idea to the instanton counting 
in the presence of surface operators.
In this way string dualities verify the observation 
that the open topological string amplitude for the toric Calabi-Yau which engineers a gauge theory
leads to the ramified instanton partition function.

From this stringy realization of surface operator,
we can also explain why the insertion of the degenerate field $\Phi_{2,1}$ 
leads to a ramified instanton partition function.
Recall the discussion of \cite{Dijkgraaf:2009pc}:
the collective fields for the matrix corresponding to the B-model
imply the Liouville field.
The mirror of $N$ toric A-brane is $N$ B-brane whose matrix description is the insertion of
\begin{align}
V_N(z)=\det (\Phi -z)^{\beta N}=e^{-N\epsilon_1\phi(z)/2},
\end{align}
where $\Phi$ is the matrix field, $\phi$ is the Liouville field and $\beta=\sqrt{-\epsilon_2/\epsilon_1}$ 
\cite{Dijkgraaf:2009pc}.
Here we use the CFT description of matrix model
$\sqrt{-\epsilon_1\epsilon_2}\partial\phi=\tr (z-\Phi)^{-1}$.
A single brane $N=1$ thus leads to the degenerate field $\Phi_{2,1}=e^{-b\phi}$ of interest.
In this way we can show that the degenerate field leads to the surface operator via the string dualities.
See \cite{Mironov:2010zs,Itoyama:2010ki,Itoyama:2009sc,Eguchi:2009gf} 
for details of the $\beta$-ensemble which is the B-model
in the generic $\Omega$-background.

In the next section, we study the AGT relation by using this topological string description of the surface operators.
We employ the refined topological vertex in order to deal with the generic $\Omega$-background $\epsilon_{1,2}$.

\section{Geometric Transition and AGT relation}
In this section we explain why
the degenerate field insertion gives the instanton partition function
in the presence of a surface operator.
We employ the concept of the geometric transition \cite{Gopakumar:1998ii,Gopakumar:1998ki,Gopakumar:1998jq}
for the purpose of studying this aspect of the AGT relation.
\subsection{Geometric engineering of surface operators}

Let us recall the two dual constructions of surface operators:
the insertion of degenerate fields in the CFT side 
and the open string version of the geometric engineering with toric branes of topological strings.
By comparing these two approachs, we reveal an interesting aspect of the AGT relation which is a
gauge theory version of the open/closed duality.

\subsubsection*{Surface operators from degenerate AGT relations}
By using the AGT relation,
we can compute an instanton partition function from the corresponding conformal block of the Liouville CFT.
The Nekrasov partition function in the presence of a surface operator also can be calculated
by introducing an additional degenerate field in the corresponding conformal block. 
\begin{figure}[htbp]
 \begin{center}
  \includegraphics[width=80mm,clip]{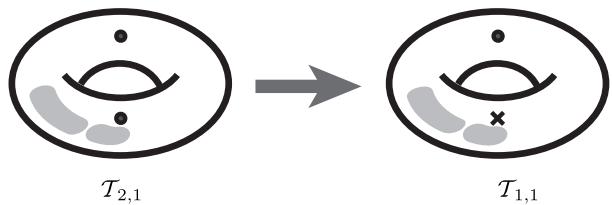}
 \end{center}
 \caption{$\mathcal{T}_{2,1}$ v.s. $\mathcal{T}_{1,1}$ : 
 torus with two punctures and torus with a puncture and a degenerate field insertion.
 The X denotes the insertion of a degenerate field.}
 \label{fig:T21vsT11}
\end{figure}
This result implies the following characteristic:
by the specialization of parameters which corresponds to change a primary to a degenerate field in the Liouville CFT side,
an ordinary Nekrasov partition function leads to a surface operator partition function of another gauge theory.
See Fig.\ref{fig:T21vsT11} for a typical example.
The torus with two primary fields (punctures) corresponding to the so-called $\hat{A}_1$ quiver gauge theory 
(a.k.a. the $\mathcal{T}_{2,1}$ theory).
On the one hand, the torus with a primary and a degenerate fields, 
which is the right hand side of Fig.\ref{fig:T21vsT11},
describes the $\mathcal{N}=2^*$ theory with an elementary surface operator.
We can therefore obtain the surface operator of $\mathcal{N}=2^*$ theory
by adjusting the parameters of the Nekrasov partition function of the $\hat{A}_1$ quiver to
the insertion of the degenerate field.

It is very unclear why surface operators emerge only by adjusting physical parameters of a gauge theory. 
In this section we explain this phenomena from the perspective of the open/closed duality.
Before studying it, we mention to the stringy version of this characteristic.

\subsubsection*{Surface operators from open topological strings}
As we saw in the previous section, 
surface operators emerge from the topological A-branes
via the geometric engineering.
Meanwhile we can compute a partition function for surface operator from a Nekrasov partition function
by choosing degenerate parameters of a gauge theory.
The open string amplitude which engineers a surface operator
therefore comes of an closed string partition function by adjusting parameters.
\begin{figure}[htbp]
 \begin{center}
  \includegraphics[width=120mm,clip]{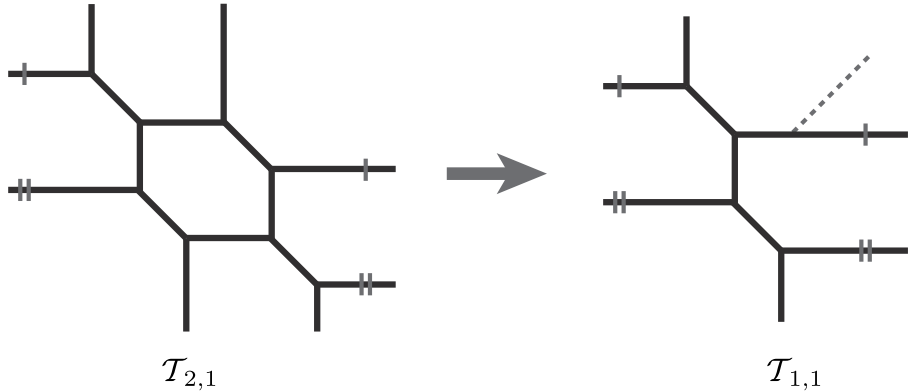}
 \end{center}
 \caption{The toric geometry which engineers the $SU(2)\times SU(2)$ 
 $\hat{A}_1$ quiver gauge theory $\mathcal{T}_{2,1}$
 and brane insertion in the $\mathcal{T}_{1,1}$ geometry.}
 \label{fig:T21toric}
\end{figure}
This stringy situation corresponding to the $\mathcal{N}=2^*$ theory 
(Fig.\ref{fig:T21vsT11}) is illustrated in Fig.\ref{fig:T21toric}.
Then a question arises: why do the open strings
be recovered from the closed strings by choosing degenerate parameters?
We will give an answer to this question by utilizing the open/closed duality,
or the geometric transition,
of the topological string theory.
\subsection{$\mathcal{T}_{4,0}$ : $SU(2)$ gauge theory with 4 flavors}

Let us consider the simple example,
$SU(2)$ gauge theory with 4 flavors (a.k.a the $\mathcal{T}_{4,0}[A_1]$ theory).
This theory is related to the sphere with four functures through the M5-brane construction.
In order to introduce a surface operator,
we need to add a degenerate field in the CFT side.
We then obtain the sphere with five functures, 
which corresponds to  the $\mathcal{T}_{5,0}[A_1]$ theory.
We therefore study the geometric engineering of the $\mathcal{T}_{5,0}[A_1]$ theory.

\subsubsection*{$\mathcal{T}_{\textbf{5,0}}$ : $\textbf{SU(2)}\times\textbf{SU(2)}$ quiver gauge theory}
\begin{figure}[htbp]
 \begin{center}
  \includegraphics[width=90mm,clip]{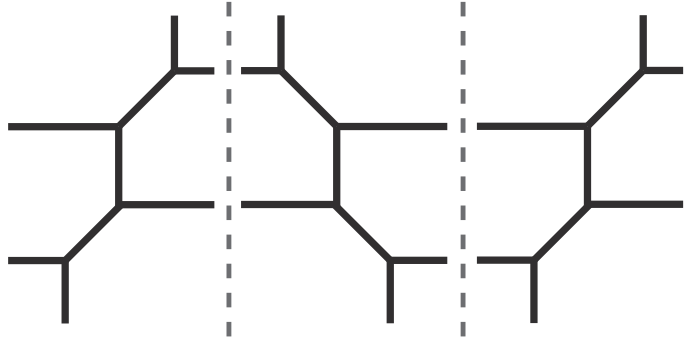}
 \end{center}
 \caption{The toric geometry which engineers the $SU(2)\times SU(2)$ 
 linear quiver gauge theory $\mathcal{T}_{5,0}$.}
 \label{fig:T50toric}
\end{figure}

The $\mathcal{T}_{{5,0}}[A_1]$ theory is the ${SU(2)}\times{SU(2)}$ quiver gauge theory
with a bifundamental hypermultiplets and two fundamentals for each $SU(2)$.
This theory is engineered by the toric Calabi-Yau geometry which is illustrated in Fig.\ref{fig:T50toric}.

The topological string partition function for the generic $\Omega$-background 
is computed by using the refined topological vertex.
We compute it by dividing this geometry into three parts in order to get the expression of the Nekrasov-type.
See Appendix B for the definition of the refined topological vertex.
We choose the horizontal direction for the preferred one of the refined vertex as in \cite{Iqbal:2007ii,Taki:2007dh}.

Let us compute the partial amplitude in the left side of Fig.\ref{fig:T50toric}.
We attach Young diagrams $Y_{1,2}$ on the two chopped edges.
Since all the bundles over the local curve $\mathbb{CP}^1$ inside the geometry 
are $\mathcal{O}(-1)\oplus\mathcal{O}(-1)$,
the framing numbers are all zero and we need not to insert any framing factors.
The topological string amplitude is then given by
\begin{align}
{Z^{L}}_{Y_1,Y_2}(t,q;\,Q_{m_1},Q_{m_2},Q_{f_1})
&=\sum_{R_i}(-Q_{m_1})^{|R_1|}(-Q_{f_1})^{|R_2|}(-Q_{m_2})^{|R_3|}\nonumber\\
&\times C_{R_1\,\varPhi\, Y_1}(t,q)C_{{R_1}^TR_2\,\varPhi}(q,t)C_{R_3 \,{R_2}^TY_2}(t,q)C_{{R_3}^T\,\varPhi\,\varPhi}(q,t)\nonumber\\
&=\prod_{a=1}^{2}q^{\frac{||Y_a||^2}{2}}\tilde{Z}_{Y_a}(t,q)
\sum_{R_i,V,W}
S_{{R_1}^T}(\,Q_{m_1}Q_{f_1}t^{-\rho}q^{-Y_1})\nonumber\\
&\times
S_{R_1/V} (-Q_{f_1}^{-1}q^{-\rho})\,S_{R_2/V} (-\sqrt{q/t}\,Q_{f_1}t^{-\rho})\nonumber\\
&\times \rule{0pt}{4ex}
S_{{R_2}^T/W} (\sqrt{t/q}\,t^{-{Y_2}^T}q^{-\rho})\,S_{{R_3}^T/W} (t^{-\rho}q^{-Y_2})\nonumber\\
&\times \rule{0pt}{4ex}
S_{{R_3}}(-Q_{m_2}q^{-\rho}).
\end{align}
This is a typical strip geometry partition function
which is ubiquitous in the geometric engineering of the superconfomal gauge theories. 
We utilize the refined vertex on a strip \cite{Iqbal:2004ne,Taki:2007dh} 
for evaluating such a summation of the Schur functions.
We thus obtain the partial amplitude for the left side of Fig.\ref{fig:T50toric}:
\begin{align}
{Z^{L}}_{Y_1,Y_2}(t,q;\,Q_{m_1},Q_{m_2},Q_{f_1})
&=\prod_{a=1}^{2}q^{\frac{||Y_a||^2}{2}}\tilde{Z}_{Y_a}(t,q)\nonumber\\
&\times\prod_{i,j=1}^\infty
\frac{(1-Q_{m_1}t^{i-1/2}q^{-Y_{1i}+j-1/2})(1-Q_{m_1}Q_{f_1}Q_{m_2}t^{j-1/2}q^{-Y_{1j}+i-1/2})}
{(1-Q_{m_1}Q_{f_1}t^{-Y_{2i}^T+j}q^{-Y_{1j}+i-1})}\nonumber\\
&\times\prod_{i,j=1}^\infty
\frac{(1-Q_{f_1}t^{-Y_{2j}^T+i-1/2}q^{j-1/2})(1-Q_{m_2}t^{i-1/2}q^{-Y_{2i}+j-1/2})}
{(1-Q_{f_1}Q_{m_2}t^{i-1}q^{j})},
\end{align}
where the K\"aher parameters are identified with the gauge theory parameters through
\begin{align}
&Q_{12}:=Q_{m_1}Q_{f_1}=e^{-2Ra},\\
&Q_{m_1}=e^{-R(a-m_1)}\,q^{1/2}t^{-1/2},\\
&Q_{m_2}=e^{-R(-a-m_2)}\,q^{1/2}t^{-1/2}.
\end{align}
Here $a$ is the Coulomb moduli parameter of the gauge factor
corresponding to the left side of the toric blocks
and $m_{1,2}$ are the masses of fundamentals for the $SU(2)$ factor.

The partial amplitude for the middle of Fig.\ref{fig:T50toric} is also a typical strip amplitude.
We can write this amplitude with the refined vertex as
\begin{align}
{Z^{M}}_{Y_1,Y_2,W_1,W_2}(t,q)
&=\sum_{R_i}(-\hat{Q}_{m_1})^{|R_1|}(-\hat{Q}_{f_1})^{|R_2|}(-\hat{Q}_{m_2})^{|R_3|}\nonumber\\
&\times C_{\varPhi\,R_1\, {Y_1}^T}(q,t)C_{R_2\,{R_1}^T{W_2}^T}(t,q)
C_{{R_2}^TR_3\,{Y_2}^T}(q,t)C_{\varPhi\,{R_3}^T{W_1}^T}(q,t)\nonumber\\
&=\prod_{a=1}^{2}t^{\frac{||Y_a^T||^2}{2}}\tilde{Z}_{Y_a^T}(q,t)\,
q^{\frac{||Y_a^T||^2}{2}}\tilde{Z}_{W_a^T}(t,q)
\sum_{R_i,V,W}
S_{R_1}\left(\,\hat{Q}_{m_1}\hat{Q}_{f_1}\sqrt{q/t}\,t^{-\rho}q^{-Y_1}\right)\nonumber\\
&\times
S_{{R_1}^T/V} \left(-\hat{Q}_{f_1}^{-1}\sqrt{t/q}\,t^{-W_2}q^{-\rho}\right)\,
S_{{R_2}^T/V} \left(-\hat{Q}_{f_1}t^{-\rho}q^{-{W_2}^T}\right)\nonumber\\
&\times \rule{0pt}{4ex}
S_{{R_2}/W} \left(t^{-{Y_2}^T}q^{-\rho}\right)\,
S_{{R_3}/W} \left( \sqrt{q/t}\,t^{-\rho}q^{-Y_2}\right)\nonumber\\
&\times \rule{0pt}{4ex}
S_{{R_3}^T}\left(-\hat{Q}_{m_2}\sqrt{t/q}\,t^{-W_1}q^{-\rho}\right).
\end{align}
Taking the summation over the Young diagrams, we obtain the following expression for the amplitude:
\begin{align}
{Z^{M}}_{Y_1,Y_2,W_1,W_2}(t,q)
&=\prod_{a=1}^{2}t^{\frac{||Y_a^T||^2}{2}}\tilde{Z}_{Y_a^T}(q,t)\,
q^{\frac{||Y_a^T||^2}{2}}\tilde{Z}_{W_a^T}(t,q)
\nonumber\\
&\times\prod_{i,j=1}^\infty
\frac{(1-\hat{Q}_{m_1}t^{-W_{2j}+i-1/2}q^{-Y_{1i}+j-1/2})
(1-\hat{Q}_{m_1}\hat{Q}_{f_1}\hat{Q}_{m_2}t^{-W_{1j}+i-1/2}q^{-Y_{1i}^T+j-1/2})}
{(1-\hat{Q}_{m_1}\hat{Q}_{f_1}t^{-Y_{2i}^T+j-1}q^{-Y_{1j}+i})}\nonumber\\
&\times\prod_{i,j=1}^\infty
\frac{(1-\hat{Q}_{f_1}q^{-W_{2i}^T+j-1/2}t^{-Y_{2j}^T+i-1/2})
(1-\hat{Q}_{m_2}t^{-W_{1j}+i-1/2}q^{-Y_{2i}+j-1/2})}
{(1-\hat{Q}_{f_1}\hat{Q}_{m_2}t^{-W_{1j}+i}q^{-W_{2i}^T+j-1})}.
\end{align}
We rewrite the K\"aher parameters as
\begin{align}
&\hat{Q}_{12}:=\hat{Q}_{m_1}\hat{Q}_{f_1}=e^{-2Ra},\\
&\hat{Q}_{m_1}=e^{-R(a-\tilde{a}-m)}q^{-1/2}t^{1/2},\\
&\hat{Q}_{m_2}=e^{-R(-a+\tilde{a}-m)}q^{-1/2}t^{1/2},
\end{align}
where $m$ is the mass of the bifundamental matter and
$\tilde{a}$ is the Coulomb moduli parameter of the gauge factor
corresponding to the right side of the toric diagram.
This factor $Z^M$ therefore gives the instanton measure for the bifundamental via the geometric enginnering.

The right side of the partition function is essentially the same as $Z^{L}$, since the refined vertex implies
\begin{align}
{Z^{R}}_{W_1,W_2}(t,q;\,\tilde{Q}_{m_1},\tilde{Q}_{m_2},\tilde{Q}_{f_1})
&=\sum_{R_i}(-\tilde{Q}_{m_1})^{|R_1|}(-\tilde{Q}_{f_1})^{|R_2|}(-\tilde{Q}_{m_2})^{|R_3|}\nonumber\\
&\times C_{R_1\,\varPhi\, W_1}(q,t)C_{{R_1}^TR_2\,\varPhi}(t,q)C_{R_3 \,{R_2}^TW_2}(q,t)C_{{R_3}^T\,\varPhi\,\varPhi}(t,q),
\end{align}
where
\begin{align}
&\tilde{Q}_{12}:=\tilde{Q}_{m_1}\tilde{Q}_{f_1}=e^{-2R\tilde{a}},\\
&\tilde{Q}_{m_1}=e^{-R(\tilde{a}+\tilde{m_1})}q^{-1/2}t^{1/2},\\
&\tilde{Q}_{m_2}=e^{-R(-\tilde{a}+\tilde{m_2})}q^{-1/2}t^{1/2}.
\end{align}
Then we can rewrite it into $Z^{L}$ as follows:
\begin{align}
{Z^{R}}_{W_1,W_2}\left(t,q;\,\tilde{a},\tilde{m_1},\tilde{m_2}\right)
=
{Z^{L}}_{W_1,W_2}\left(q,t;\,
\tilde{a},-\tilde{m_1},-\tilde{m_2}\right).
\end{align}

The full partition function is obtained by gluing these three partial amplitudes.
By using formulae of Appendix B, we can recast it into the instanton measures.
Then the normalized partition function is precisely the Nekrasov partition function for the 
 $\mathcal{T}_{{5,0}}[A_1]$ theory:
\begin{align}
&Z_{\,\mathcal{T}_{5,0},\, 5D}\nonumber\\
&=\sum_{Y_{1,2},\,W_{1,2}}
Q^{|Y_1|+|Y_2|}\tilde{Q}^{|W_1|+|W_2|}
\frac{
{Z^{L}}_{Y_1,Y_2} (t,q;\, a,\, m_1,m_2)
{Z^{M}}_{Y_1,Y_2,W_2^T,W_1^T}(t,q)
{Z^{L}}_{W_2^T,W_1^T} \left(q,t;\,\tilde{a},\tilde{m_1},\tilde{m_2}\right)
}{
{Z^{L}}_{\varPhi,\,\varPhi}\left(t,q;\,{a},\,{m_1},m_2\right)
{Z^{M}}_{\varPhi,\,\varPhi,\,\varPhi,\,\varPhi}(t,q)
{Z^{L}}_{\varPhi,\,\varPhi}\left(q,t;\,\tilde{a},-\tilde{m_1},-\tilde{m_2}\right)
}.\nonumber\\
&=\sum_{Y_{1,2},\,W_{1,2}}
M\,Q^{|Y_1|+|Y_2|}\tilde{Q}^{|W_1|+|W_2|}
z_{\textrm{vect.}, 5D}(\vec{a},\vec{Y};t,q)
\prod_{f=1,2}z_{\textrm{fund.}, 5D}(\vec{a},\vec{Y},m_f;t,q)
\nonumber\\
&\qquad\times z_{\textrm{bifund.}, 5D}(\vec{a},\vec{Y};\vec{\tilde{a}},\vec{W};m;t,q)
z_{\textrm{vect.}, 5D}(\vec{\tilde{a}},\vec{W};t,q)
\prod_{f=1,2}z_{\textrm{fund.}, 5D}(\vec{\tilde{a}},\vec{W},\tilde{m}_f;t,q),
\end{align}
where $M$ is a monomial of $q$, $t$ and $Q_*$, which is irrelevant
when taking the four dimensional limit.
In fact, by taking the 4-dimensional field theory limit,
this topological string partition function recovers the Nekrasov
instanton 
partition function.
\subsection*{Degenerate K\"aher moduli and bubbling}
Now let us move on to the emergence of a surface operator.
The AGT relation suggests that the following specialization
which corresponds to the degeneration of a primary field
leads to the Nekrasov partition function in the presence of a surface operator:
\begin{align}
\tilde{a}=-\tilde{m}_1=\tilde{m}_2+\epsilon_1.
\end{align}
Then the K\"aher parameters of the toric geometry take the quantized values
\begin{align}
\label{Nf4dK1}
&\tilde{Q}_{m_1}=q^{-3/2}t^{1/2}\,(\,\to q^{-1}\quad \textrm{when} \quad q=t\,),\\
\label{Nf4dK2}
&\tilde{Q}_{m_2}=q^{-1/2}t^{1/2}\,(\,\to 1\quad \textrm{when} \quad q=t\,).
\end{align}
Note that $\tilde{t}_{m_1}$ and $\tilde{t}_{m_2}$ here are essentially $0$ and $g_s$,
where $g_s$ is the topological string coupling constant.
The geometric transition will explain this quantization of the K\"aher parameters.

\begin{figure}[htbp]
 \begin{center}
  \includegraphics[width=90mm,clip]{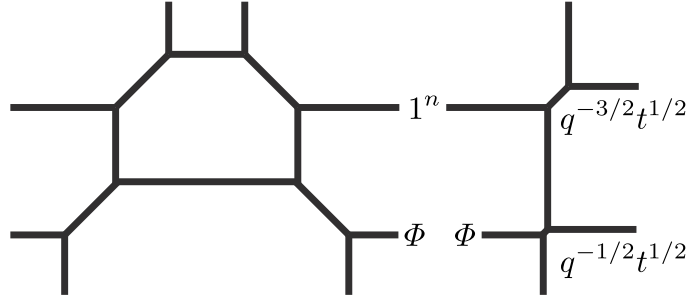}
 \end{center}
 \caption{The toric geometry which engineers the $SU(2)$ gauge theory with four flavors
in the presence of the simple surface operator.}
 \label{fig:quiver40surface1}
\end{figure}

This degeneration of the K\"aher parameters simplifies
the topological string partition function.
Let us focus on the instanton measure coming from the fundamentals:
\begin{align}
\prod_{f=1,2}z_{\textrm{fund.}, 5D}(\vec{\tilde{a}},\vec{W},\tilde{m}_f;t,q)
\to
\prod_{(i,j)\in W_1}(1-t^{i-1}q^{-j+2})
\prod_{(i,j)\in W_2}(1-t^{i-1}q^{-j+1})
\end{align}
From these factors, 
the Young diagram $W_1$ must not contain the box $(1,2)$ and 
$W_2$ must be the empty Young diagram $\varPhi$
in order to give a non-zero contribution to the partition function.
This means that the Young diagrams $\vec{W}$ have to take the following form:
\begin{align}
W_1=[1^n] \,\,\,(\, n=0,1,2,\cdots\,),\quad W_2=\varPhi.
\end{align}
These degenerate parameters thus cut off the toric geometry of the original theory as Fig.\ref{fig:quiver40surface1}.
This toric geometry is very resemble to that of the $SU(2)$ gauge theory with four flavors.
The reason of it will soon be clear.

The limitation $\vec{W}=([1^n], \varPhi)$ simplifies 
the partition function for  the $\mathcal{T}_{{5,0}}[A_1]$ theory.
Due to the AGT relation, 
the resulting partition function is precisely the ramified instanton partition function for the $N_f=4$ theory.
In this way the degeneration of parameters gives the five-dimensional Nekrasov partition
for  the $SU(2)$ gauge theory with four flavors in the presence of a surface operator:
\begin{align}
\Psi_{\,\mathcal{T}_{4,0}, 5D}(z)
&=\sum_{Y_{1,2},\,n}
z^{n}Q^{|Y_1|+|Y_2|}
z_{\textrm{vect.}, 5D}(\vec{a},\vec{Y};t,q)
\prod_{f=1,2}z_{\textrm{fund.}, 5D}(\vec{a},\vec{Y},m_f;t,q)
\nonumber\\
&\qquad\times z_{\textrm{bifund.}, 5D}(\vec{a},\vec{Y};(-\tilde{m}_1,\,\tilde{m}_1),([1^n],\varPhi);m;t,q)
\nonumber\\
&\qquad \times \rule{0pt}{4ex}
z_{\textrm{vect.}, 5D}((-\tilde{m}_1,\,\tilde{m}_1),([1^n],\varPhi);t,q)
\prod_{i=1}^n(1-e^{2R\tilde{m}_1}t^{i-1})(1-t^{i-1}q).
\end{align}
We can recast it into the Nekrasov partition function for the $N_f=4$ theory
with an added insertion of the factor coming of the surface operator
\begin{align}
\label{Nf4surf}
\Psi_{\,\mathcal{T}_{4,0}, 5D}(z)
&=\sum_{Y_{1,2},\,n}
z^{n}Q^{|Y_1|+|Y_2|}
z_{\textrm{vect.}, 5D}(\vec{a},\vec{Y};t,q)
\prod_{f=1}^4z_{\textrm{fund.}, 5D}(\vec{a},\vec{Y},m_f;t,q)
\nonumber\\
&\qquad\times 
\prod_{a=1}^2 \frac{N_{[1^n],Y_a} (e^{-R(a_a-m_4+\epsilon)};q;t)}
{N_{\varPhi ,Y_a} (e^{-R(a_a-m_4+\epsilon)};q;t)}
\prod_{i=1}^n
\frac{1}{(1-t^i)(1-e^{-R(m_3-m_4)}t^i q^{-1})},
\end{align}
where we introduce the mass parameters
\begin{align}
m_3=m-\tilde{m}_1+\epsilon,\quad m_4=m+\tilde{m}_1+\epsilon.
\end{align}
This formula is the five-dimensional lift of the result of \cite{Kozcaz:2010af}.
While the four-dimensional version of (\ref{Nf4surf}) 
was derived from the conformal block via the AGT relation in \cite{Kozcaz:2010af}, 
we employ the topological vertex for engineering the ramified instanton partition function.

\begin{figure}[htbp]
 \begin{center}
  \includegraphics[width=130mm,clip]{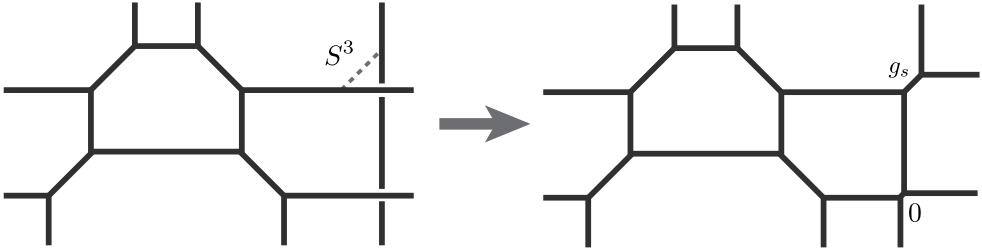}
 \end{center}
 \caption{The geometric transition relates the surface-operator/toric-brane insertion
and the bubbling geometry which engineers the $SU(2)\times SU(2)$ 
linear quiver gauge theory $\mathcal{T}_{5,0}$.}
\label{fig:bubb1}
\end{figure}

Now let us move on to make the meaning of the specialization (\ref{Nf4dK1}) and (\ref{Nf4dK2}) clear.
The point is that the toric brane realization of surface operators leads to
these special values through the open/closed duality.
Let us consider the system which is drawn on the left side of Fig.\ref{fig:bubb1}.
As we have discussed in the previous section,
the Lagrangian brane denoted by the broken line in Fig.\ref{fig:bubb1}
engineers the surface operator.
Here we introduce the framing in order to compactify the Lagrangian cycle.
Such a compactification does not essentially change the amplitude since the A-model topological string theory
is independent of the complex structure moduli.
Then we apply the geometric transition \cite{Gopakumar:1998ii} 
which changes branes into the background geometry.
The multiplicity of branes then becomes the K\"ahler parameter of the emerged cycle.
The geometric transition for our system is shown in Fig.\ref{fig:bubb1}.
Since we have inserted only one brane on the upper leg as in Fig.\ref{fig:bubb1},
a new $\mathbb{CP}^1$ whose size is $\tilde{t}_1=g_s$ emerges into the toric geometry after the transition.
There is also the Lagrangian cycle $S^3$ without a brane.
This Lagrangian three-cycle gives the zero-size  $\mathbb{CP}^1$ through the transition.
In this way, the geometric transition relates the two reaization of the surface operator:
the open topological string approach and the closed topological string for degenerate parameters.
The refined version of the brane-induced parameters $\tilde{t}_{1,2}=g_s, 0$ is precisely equal to
(\ref{Nf4dK1}) and (\ref{Nf4dK2}) which were derived from the CFT side.

The phenomenon which changes branes into two-cycles is known as the bubbling Calabi-Yau 
\cite{Gomis:2006mv,Gomis:2007kz}.
The field theory limit of such a duality then proves the implication of the AGT relation:
the ramified instanton partition function for an elementary surface operator
emerges from the Nekrasov partition function by limiting the parameters.
This feature provides an efficient tool to compute ramified instanton partition functions.

\subsection{Generalization to  $SU(N_c)$ gauge theory with $2N_c$ flavors}
\begin{figure}[tbp]
 \begin{center}
  \includegraphics[width=80mm,clip]{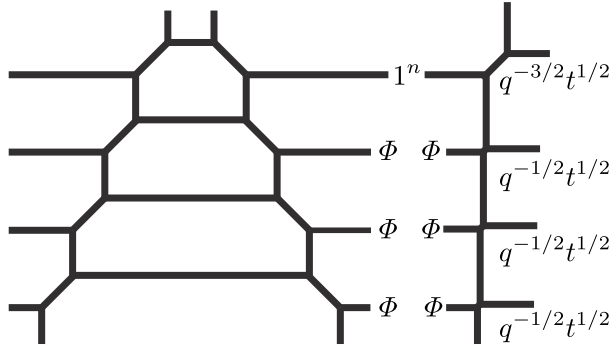}
 \end{center}
 \caption{The toric geometry which engineers the $SU(N_c)$ $N_f=2N_c$ gauge theory
in the presence of the elementary surface operator.}
 \label{fig:quiver40surface2}
\end{figure}

We can easily generalize this result to
the $\mathcal{T}_{{4,0}}[A_{N_c-1}]$ theory.
The toric geometry for the $SU(N_c)$ gauge theory
can be obtained by gluing the geometries for $SU(2)$ theory in vertical direction.
See \cite{Iqbal:2004ne} for details.
We choose the degenerate parameters as in Fig.\ref{fig:quiver40surface2}.
The refined vertex on the strip geometry then leads to the following factors 
for each $\mathcal{O}(-1)\oplus\mathcal{O}(-1)$ curve inside the strip geometry
\begin{align}
\prod_{(i,j)\in W_1}(1-t^{i-1}q^{-j+2})\prod_{a=2}^{N_c}\prod_{(i,j)\in W_a}(1-t^{i-1}q^{-j+1}).
\end{align}
The Young diagrams take the following form give the non-zero contribution for the partition function:
\begin{align}
W_1=[1^n] \,\,\,(\, n=0,1,2,\cdots\,),\quad W_{2,3,\cdots,N_c}=\varPhi.
\end{align}
Therefore the toric geometry Fig.\ref{fig:quiver40surface2} 
engineers the ramified instanton partition function
for the $SU(N_c)$ gauge theory with $2N_c$ flavors.
\subsection{$\mathcal{T}_{1,1}$ : $\mathcal{N}=2^{*}$ $SU(2)$ gauge theory}
\begin{figure}[tbp]
 \begin{center}
  \includegraphics[width=40mm,clip]{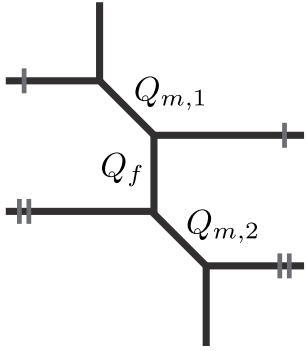}
 \end{center}
 \caption{The toric geometry which engineers the $\mathcal{N} =2^*$ gauge theory $\mathcal{T}_{1,1}$.
 The horizontal direction is periodic by the identification of edges.}
 \label{fig;T11}
\end{figure}

Let us next consider the $\mathcal{N}=2^{*}$ $SU(2)$ gauge theory:
the $SU(2)$ gauge theory with an adjoint matter.
The toric geometry for the theory is illustrated in Fig.\ref{fig;T11}.
The refined vertex on the geometry leads to the Nekrasov partition function
for the $\mathcal{N}=2^{*}$ $SU(2)$ gauge theory:
\begin{align}
Z_{\mathcal{N}=2^{*}, 5D}
=\sum_{Y_{1,2}}
Q^{|Y_1|+|Y_2|}
\frac{{Z^{M}}_{Y_1,\,Y_2,\,{Y_2}^T,{Y_1}^T}(t,q)}
{{Z^{M}}_{\varPhi,\,\varPhi,\,\varPhi,\,\varPhi}(t,q)}.
\end{align}
It is easy to rewrite it into the standard form of the Nekrasov partition function
by using the formulae in Appendix B.
Here the identification between the K\"ahler parameters 
and the gauge theory parameters is 
\begin{align}
&Q_{f_1}Q_{m_1}=e^{-2Ra}q^{-1/2}t^{1/2},\\
&Q_{m_1}=Q_{m_2}=e^{Rm}q^{-1/2}t^{1/2}.
\end{align}
See \cite{Hollowood:2003cv} for the detailed study of such a partition function.

\subsubsection*{$\mathcal{T}_{\textbf{2,1}}$ : $\hat{A}_1$ quiver gauge theory}
\begin{figure}[htbp]
 \begin{center}
  \includegraphics[width=60mm,clip]{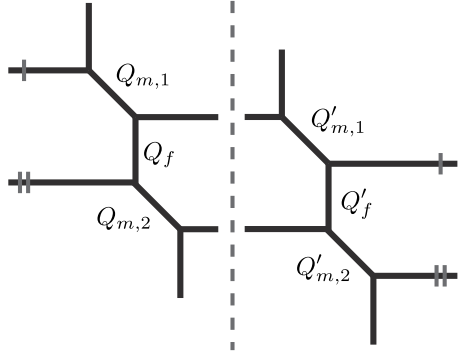}
 \end{center}
 \caption{The toric geometry which engineers the $\hat{A}_1$ gauge theory $\mathcal{T}_{2,1}$
 .}
 \label{fig;T21}
\end{figure}

As we have mentioned at the beginning of this section,
the $\hat{A}_1$ quiver gauge theory leads to
the surface operator of the $\mathcal{N}=2^{*}$ theory
via the AGT relation.
This quiver gauge theory is also realized  by the string theory on the toric Calabi-Yau Fig.\ref{fig;T21}.
The refined vertex on the geometry implies the following partition function:
\begin{align}
Z_{\hat{A}_1, 5D}=\sum_{Y_{1,2},\,W_{1,2}}
Q^{|Y_1|+|Y_2|}
\tilde{Q}^{|W_1|+|W_2|}
\frac{{Z^{M}}_{Y_1,Y_2,{W_2}^T,{W_1}^T}(t,q)\,
{Z^{M}}_{{W_1},{W_2},{Y_2}^T,{Y_1}^T}(q,t)}
{{Z^{M}}_{\varPhi,\,\varPhi,\,\varPhi,\,\varPhi}(t,q)\,
{Z^{M}}_{\varPhi,\,\varPhi,\,\varPhi,\,\varPhi}(q,t)}.
\end{align}
By using the formulae in Appendix B, we can show easily that 
this is equal to the five-dimensional Nekrasov partition function
for $\hat{A}_1$ theory
\begin{align}
\label{A1hat}
Z_{\hat{A}_1, 5D}&=\sum_{Y_{1,2},\,W_{1,2}}
M\,Q^{|Y_1|+|Y_2|}\,{Q^\prime}^{|W_1|+|W_2|}
z_{\textrm{vect}, 5D}(\vec{a},\vec{Y})
z_{\textrm{vect}, 5D}(\vec{b},\vec{W})
\nonumber\\
&\qquad\qquad\qquad\qquad \times z_{\textrm{bifund}, 5D}(\vec{a},\vec{Y};\vec{b},\vec{W};m)
z_{\textrm{bifund}, 5D}(\vec{b},\vec{W};\vec{a},\vec{Y};m^\prime).
\end{align}
Here $M$ is a monomial which is irrelevant in the four-dimensional limit.
Now let us verify our expectation that the ramified instanton partition function for the $\mathcal{N}=2^{*}$ theory
is equal to the specialization of (\ref{A1hat}).
Before we turn to the derivation from the string theory,
we study the Liouville CFT side to find the corrsponding degeneration of parameters.

\subsubsection*{Conformal block for $\mathcal{T}_{\textbf{2,1}}$ theory and its degeneration}
\begin{figure}[htbp]
 \begin{center}
  \includegraphics[width=40mm,clip]{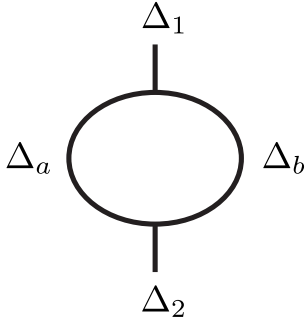}
 \end{center}
 \caption{The conformal block for torus with 2 points, which corresponds to
 $SU(2)$ $\hat{A}_{1}$-quiver gauge theory.}
\label{fig;cbtorus2}
\end{figure}

Recall the AGT relation for the $\mathcal{T}_{{2,1}}$ theory: 
the conformal block \cite{Marshakov:2009gs,Mironov:2009dr,Alba:2009fp,Alba:2009ya}
for torus with two punctures Fig.\ref{fig;cbtorus2}
is the counterpart of the Nekrasov partition function for the $\hat{A}_1$ gauge theory.
We can compute the conformal block perturbatively as
\begin{align}
&\mathcal{B}_{\,\mathcal{T}_{{2,1}}}(q_1,q_2)
=\nonumber\\
&1+
q_1\frac{(\Delta_1-\Delta_a-\Delta_b)(\Delta_2+\Delta_a-\Delta_b)}{2\Delta_a}
+q_2\frac{(\Delta_1-\Delta_b-\Delta_a)(\Delta_2+\Delta_b-\Delta_a)}{2\Delta_b}
+\cdots.
\end{align}
The masses of the $\hat{A}_1$ theory relate to the external Liouville momentum 
of the conformal block $\epsilon_1\epsilon_2\Delta_{i}=m_i(\epsilon-m_i)$.
The $U(1)$ part of the Nekrasov partition function,
which is a crucial factor for the gauge/Liouville correspondence \cite{Alday:2009aq},
is given by
\begin{align}
Z_{\,\mathcal{T}_{{2,1}}}^{U(1)}(q_1,q_2)
&=
\prod_{l=0}^\infty
(1-q_1^{l+1}q_2^{l+1})^{(2m_1(\epsilon-m_1)+2m_2(\epsilon-m_2))/\epsilon_1\epsilon_2-1}\nonumber\\
&\qquad\qquad\qquad\quad\times
(1-q_1^{l+1}q_2^{l})^{2m_2(\epsilon-m_1)/\epsilon_1\epsilon_2}
(1-q_1^{l}q_2^{l+1})^{2m_1(\epsilon-m_2)/\epsilon_1\epsilon_2}\nonumber\\
&=
1-q_1\frac{2m_2(\epsilon-m_1)}{\epsilon_1\epsilon_2}
-q_2\frac{2m_1(\epsilon-m_2)}{\epsilon_1\epsilon_2}
+\cdots.
\end{align}
From the perspective of the AGT relation,
this factor comes of the free-field computation of the corresponding conformal block \cite{Marshakov:2009gs}.
Based on explicit computation, we can confirm that 
the product of these factors $Z^{U(1)}\cdot\mathcal{B}$
is equal to
the Nekrasov partition function for the $SU(2)$  $\hat{A}_1$ gauge theory  \cite{Alday:2009aq}.

Let us consider the degeneration of the external vertex operator  $V_{m_2}(q_2)\to \Phi_{2,1}(q_2)$.
The fusion rule then implies the following specialization of parameters:
\begin{align}
\Delta_a=\frac{\epsilon^2}{4}-a^2,\quad
\Delta_b=\frac{\epsilon^2}{4}-\left(a+\frac{\epsilon_2}{2} \right)^2,\quad
\Delta_2=-\frac{\epsilon_1\epsilon_2}{2}-\frac{3\epsilon_2^2}{4}
\end{align}
Notice that this situation implies the following specialization of parameters of the gauge theory
\begin{align}
\label{degparamT21}
m_2=-\frac{\epsilon_2}{2},\quad
b=a+\frac{\epsilon_2}{2}.
\end{align}
Then the degenerated conformal block becomes
\begin{align}
\mathcal{B}_{\,\mathcal{T}_{{2,1}}}(q_1,q_2)
&=1
-q_1\frac{a\epsilon_2+\epsilon_2^2/4-m_1^2+m_1\epsilon}{\epsilon_1(2a+\epsilon)}
+q_2\frac{-a\epsilon_2-\epsilon_2^2/4-m_1^2+m_1\epsilon}{\epsilon_1(2a-\epsilon_1)}
+\cdots.\\
Z_{\,\mathcal{T}_{{2,1}}}^{U(1)}(q_1,q_2)
&=
1-q_1\frac{(\epsilon-m_1)}{\epsilon_1}
-q_2\frac{m_1(2\epsilon+\epsilon_2)}{\epsilon_1\epsilon_2}
+\cdots.
\end{align}
The product of them is the partition function for the surface operator 
since the degenerate field corresponds to the operator via the AGT relation.
These expressions thus lead to the ramified instanton partition function
for the $\mathcal{N}=2^*$ theory:
\begin{align}
\label{cbT11U1}
\Psi_{\,\mathcal{T}_{{1,1}}}(q_1,q_2)
&=
Z_{\,\mathcal{T}_{{2,1}}}^{U(1)}\mathcal{B}_{\,\mathcal{T}_{{2,1}}}\nonumber\\
&=1
+q_1\frac{(\epsilon-\epsilon_2/2-m_1)(2a+\epsilon+\epsilon_2/2-m_1)}{\epsilon_1(2a+\epsilon)}
+\cdots.
\end{align}
We can also recover this result by computing the corresponding refined vertex amplitude (\ref{A1hat})
for the following degenerate parameters (\ref{degparamT21}).

\subsection*{Degenerate K\"aher moduli and bubbling}
\begin{figure}[htbp]
 \begin{center}
  \includegraphics[width=120mm,clip]{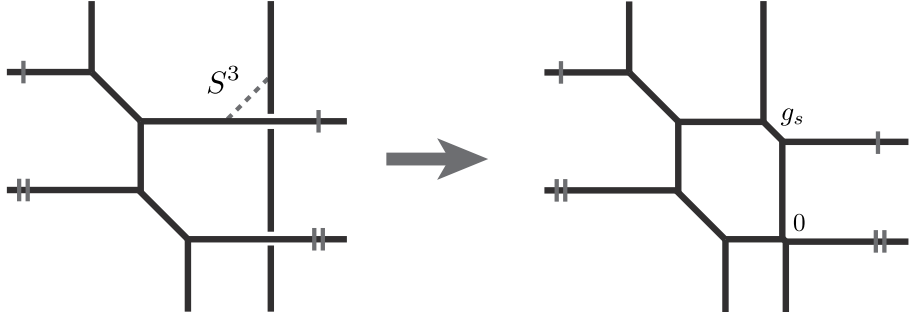}
 \end{center}
 \caption{The toric geometry which engineers the $SU(2)\times SU(2)$ 
 $\hat{A}_1$ quiver gauge theory $\mathcal{T}_{5,0}$.}
\label{fig:bubb2}
\end{figure}

Let us interpret the degenerate parameters (\ref{degparamT21}).
We saw in the case of the $N_f=4$ theory that
the toric brane construction leads to such a degenerate Calabi-Yau geometry
after the geometric transition.
This idea works also for the $\mathcal{N}=2^*$ theory.
The transition of the elliptic geometry for the $\mathcal{N}=2^*$ theory
is shown in Fig.\ref{fig:bubb2}.
From the viewpoint of the geometric engineering of the $\mathcal{T}_{{2,1}}[A_1]$ theory,
the degenerated parameters (\ref{degparamT21}) correspond to the special choice of
the K\"ahler parameters:
\begin{align}
Q^\prime_{m,1}=q\sqrt{\frac{q}{t}}
,\quad
Q^\prime_{m,2}=\sqrt{\frac{q}{t}}.
\end{align}
This specialization reflects the A-brane 
which was inserted in the upper leg before the transition as in Fig.\ref{fig:bubb2}. 

In this way the five-dimensional lift of the ramified instanton partition function is
\begin{align}
\label{2star}
\Psi_{\,\mathcal{N}=2^*}(q,z)
&=
\sum_{\vec{Y},\vec{W}}q_1^{|\vec{Y}|}q_2^{|\vec{W}|}
z_{\textrm{vect},5D}(a,\vec{Y})
z_{\textrm{vect},5D}(b=a+\epsilon_2/2,\vec{W})\nonumber\\
&\times
z_{\textrm{bifund},5D}(a,\vec{Y};b=a+\epsilon_2/2,\vec{W}; m)
z_{\textrm{bifund},5D}(b=a+\epsilon_2/2,\vec{W};a,\vec{Y}; m_2=-\epsilon_2/2).
\end{align}
Here $q_1=q^2z^{-1}$ and $q_2=-q^2z$.
It is easy to take the four-dimensional limit $R\to 0$.
We can see that the four-dimensional result equal to (\ref{cbT11U1}) based on the explicit computation.

\subsection{Decoupling limit and pure $SU(2)$ Yang-Mills theory}
So far we studied superconformal gauge theories.
Finally let us consider asymptotically free theory \cite{Gaiotto:2009ma,Marshakov:2009gn}.
For our result (\ref{2star}), we can take the decoupling limit of the massive hypermultiplet.
We propose the following decoupling limit
\begin{align}
m\to \infty,\quad m^2q_1\to\Lambda^4\lambda^{-1},\quad m^2q_2\to -\lambda.
\end{align}
Here $\lambda=\Lambda^2z$ is the two-dimensional instanton factor \cite{Gaiotto:2009fs,Maruyoshi:2010iu}.
Then we find the following expression for the ramified instanton partition function
of the $\mathcal{N}=2$ pure $SU(2)$ Yang-Mills theory:
\begin{align}
\label{ramifiedpure}
\Psi_{\,\textrm{pure SYM}}(q_1,q_2)
&=
\sum_{\vec{Y},\vec{W}}
{(\Lambda^4\lambda^{-1})}^{|\vec{Y}|}
(-\lambda)^{|\vec{W}|}
z_{\textrm{vect}}(a,\vec{Y})
z_{\textrm{vect}}(b=a+\epsilon_2/2,\vec{W})\nonumber\\
&\qquad\times
z_{\textrm{bifund}}(b=a+\epsilon_2/2,\vec{W};a,\vec{Y}; m=-\epsilon_2/2).
\end{align}
One can compute the four-dimensional case by replacing the instanton measures as follows:
\begin{align}
&z_{\textrm{vect}}(a,\vec{Y})
\to
\prod_{a,b=1,2}\prod_{(i,j)\in Y_a}
(a_a-a_b+\epsilon_1(-Y_{bj}^T+i)+\epsilon_2(Y_{ai}-j+1)))^{-1}
\nonumber\\
&\qquad\qquad\qquad\qquad\qquad \times \prod_{(i,j)\in Y_b}
(a_a-a_b+\epsilon_1(Y_{aj}^T-i+1)+\epsilon_2(-Y_{bi}+j)))^{-1},
\\
&z_{\textrm{bifund}}(a,\vec{Y};b,\vec{W};m)
\to
\prod_{a,b=1,2}\prod_{(i,j)\in Y_a}
(a_a-b_b-m+\epsilon_1(-W_{bj}^T+i)+\epsilon_2(Y_{ai}-j+1)))
\nonumber\\
&\qquad\qquad\qquad\qquad\qquad \times \prod_{(i,j)\in W_b}
(a_a-b_b-m+\epsilon_1(Y_{aj}^T-i+1)) +\epsilon_2(-W_{bi}+j)),
\end{align}
where $a_1=-a_2=a$ and $b_1=-b_2=b$.

Recently the author and Maruyoshi \cite{Maruyoshi:2010iu} 
proposed the irregular conformal block for the
ramified insanton partition function:
\begin{align}
\label{2d4dinstanton}
\Psi_{\,\textrm{pure SYM}}(q_1,q_2)
&=\sum_{n=0}^{\infty}\sum_{{Y}}\Lambda^{2|{Y}|+2n}\,
z^{n-|Y|}
\beta^{Y}_{1^{n}}
Q_{\Delta}^{-1}([1^{|Y|}]; Y)
\nonumber\\
&=1+\lambda\,\frac{1}{2\,\epsilon_1\left(a-\frac{\epsilon_1}{2}\right)}
+\Lambda^4 \lambda^{-1}\,
\frac{1}{2\,\epsilon_1\left(a+\frac{\epsilon_2}{2}+\frac{\epsilon_1}{2}\right)}\nonumber\\
&\quad
+ \lambda^{2}\,
\frac{1}
{8\,\epsilon_1^2\left(a-\frac{\epsilon_1}{2}\right)\left(a-\epsilon_1\right)}
+ \cdots,
\end{align}
where $\beta$ is an expansion coefficient of the degenerate field on the Verma module:
\begin{align}
\Phi_{2,1}(z)|\Delta, Y \rangle
=\sum_{Y^{\prime}}
z^{|Y^{\prime}|-|Y|+\delta}
\beta^{\,Y}_{\,Y^{\prime}}
\,|\Delta^\prime, Y^{\prime} \rangle.
\end{align}
Our formula (\ref{ramifiedpure}) in the four-dimensional limit $R\to 0$ will provide the combinatorical expression
for the representation-theoretical expression (\ref{2d4dinstanton}).
We can see the agreement of these results based on the explicit computation.
This relation between (\ref{ramifiedpure}) and (\ref{2d4dinstanton})
provides the extension of  the asymptotically free  AGT relation \cite{Gaiotto:2009ma,Marshakov:2009gn}
for surface operators.

\section{Conclusion and Discussion}
In this work we study surface operators by using topological string theory and dualities.
Our study suggests that
topological string theory provides the powerful way to reveal the mechanism
behind the mysterious AGT relation.

We study the geometric engineering of the Nekrasov partition function in the presence of the surface operator,
which we call the ramified instanton partition function.
We find that a toric A-brane corresponds to a surface operator of  gauge theory through string dualities.
We show that we can derive Gaiotto curves from the wave functions for the open topological string amplitudes
based on the idea.
This approach will be an alternative to the matrix model computation of the B-model in \cite{Kozcaz:2010af}.

We also show that the insertion of surface operators leads to the bubbling of Calabi-Yau geometry.
By using this idea, we compute the ramified instanton partition functions for the $N_f=4$, $\mathcal{N}=2^*$
and pure Yang-Mills theory.
Our approach provides an efficient way to compute the ramified instanton partition functions
for elementary surface operators and their five-dimensional lifts.

There are many open problems related to our study.
For example, it would be very interesting
to develop topological string realization for
more generic type of surface operators such as the full type in \cite{Alday:2010vg}.
Since these operators appear when we study $SU(N\geq3)$ gauge theories \cite{Mironov:2009by,Taki:2009zd},
we would get the idea from
topological strings on toric geometries with long strip.
It would be also important to study relation to the Hitchin systems \cite{Teschner,Bonelli:2009zp}, 
which are expected to play a key role in the AGT relation.
We want to come back to these problems in the future.

\section*{Acknowledgements}
  I would like to thank Kazunobu Maruyoshi
  for useful comments and discussion.
  This work is supported by JSPS Grant-in-Aid for Creative Scientific Research, No.19GS0219.
\appendix

\section*{Appendix}
\section{Young diagrams and Schur functions}

A Young diagram is a sequence of decreasing non-negative integers such that
\begin{align}
Y  = \left\{ \,Y_i  \in \mathbb{Z} \ge 0 |\, Y_1 \ge Y_2  \ge  \cdots \right\}.
\end{align}
The transpose of a Young diagram $Y $ is defined as follows
\begin{align}
Y^T  = \left\{ \, {Y_j}^T \in \mathbb{Z}_{ \ge 0} |\, {Y_j}^T  = \# \left\{ \, i|\, Y_i \ge j \right\}\right\}.
\end{align}
We defines the following quantities
\begin{align}
\left| Y  \right| = \sum\limits_{i = 1}^{d(Y )} { Y_i } ,\quad   \left\| Y  \right\|^2  = \sum_{i = 1}^{d(Y )} {Y_i^2 },
\quad { \kappa_Y = \sum_{(i,j) \in Y } {(j - i)} }.
\end{align}
Note that we have $\kappa_Y=\left\|Y\right\|^2-\left\|Y^T\right\|^2$.

The Schur functions are symmetric functions which are labeled by the Young diagrams.
It is the trace of $X=\textrm{diag} (x_1,x_2,\cdots)$ in a representation $Y$
\begin{align}
S_Y (x)=\textrm{Tr}_Y X .
\end{align}
See \cite{macdonald} for details about the symmetric functions.
We can also define the skew Schur functions $S_{Y_1/Y_2}=\sum_Y c^{Y_1}_{Y_2,Y}S_Y$
where $c$ is the coefficient $Y_2\otimes Y=\sum c^{Y_1}_{Y_2,Y} Y_1$.
These functions then satisfy the following properties
\begin{align}
&\rule{0pt}{4ex}S_Y  (Qx) =  Q^{\left| Y  \right|} S_Y  (x),\\
&\rule{0pt}{4ex}S_{Y /V } (Qx) = Q^{\left| Y  \right| - \left| V  \right|}  S_{Y /V } (x),\\
&\rule{0pt}{4ex}S_Y  ( q^\rho  ) =q^{\frac{\kappa_Y}{2}} S_{Y^T } (q^\rho  ) 
={( - 1)}^{\left| Y  \right|} Y_{ Y^T } ( q^{ - \rho } ).
\end{align}
Here $q^\rho$ denotes the variable $x_i=q^{-i+1/2}$.
The following identities, which are known as the Cauchy formulae,
 are most useful for the computation of the topological vertex
\begin{align}
&\rule{0pt}{4ex}\sum_Y  { S_Y  (x) S_Y  (y)}  = \prod_{i,j} {(1 -  x_i  y_j )^{ - 1} },\\
&\rule{0pt}{4ex}\sum_Y  { S_{Y^T } (x) s_Y  (y)}  = \prod_{i,j} {(1 +  x_i y_j )},\\
&\rule{0pt}{4ex}\sum_Y  S_{Y/V}(x) S_{Y/W}(y) = \prod_{i,j} (1 - x_i y_j )^{ - 1} 
 \sum_Y  S_{V/Y } (y) S_{ W/Y } (x),\\
&\rule{0pt}{4ex}\sum_Y  S_{Y^T /V } (x) S_{Y/W } (y) = \prod_{i,j}(1 +  x_i  y_j ) \sum_Y  S_{V/Y^T } (y) S_{W^T /Y^T } (x).
\end{align}
We make good use of these formulae in order to compute a vertex on a strip.
See \cite{Iqbal:2004ne,Taki:2007dh} for details. 

\section{Five-dimensional Nekrasov Partition Function}
\label{sec:5dNek}
\subsection{Instanton partition function}
In this section, we explain how the Nekrasov partition functions 
are derived.
The point is that an instanton partition function becomes more simpler
by taking the five dimensional gauge theory with twisted boundary condition
$(\vec{x},x_4)\sim (\exp R\Omega \cdot\vec{x}, x_4+R)$,
where $R$ is the radius of the fifth dimensional $S^1$
and $\Omega$ is a generator of the rotational symmetry of the space-time $\mathbb{R}^2\times\mathbb{R}^2$
\begin{align}
\Omega=\left(\begin{array}{cc} \epsilon_1\mathbb{J} & \textbf{0}\\ \textbf{0} & \epsilon_2\mathbb{J} \end{array}\right).
\end{align}

\subsection*{Vector multiplet}
The $k$-instanton part of the five dimensional pure super Yang-Mills theory  
reduces to the supersymmetric quantum mechanics
on the instantin moduli space $\mathcal{M}_k$ by taking small coupling limit.
The instanton partition function is then defined as the Witten index of the quantum mechanics
which is the index of the Dirac operator.
The Atiyah-Singer index theorem leads to the following expression:
\begin{align}
Z_{\textrm{pure},\, k ;\, 5D}&=\textrm{Ind}\,\mathcal{D}\nonumber\\
&=\int_{\mathcal{M}_k}
\hat{A}(T\mathcal{M}_k).
\end{align}
Here $\hat{A}$ is the A-roof Dirac genus
\begin{align}
\hat{A}(TM)=\prod_{i=1}^r\frac{Rx_i}{e^{Rx_i/2}-e^{-Rx_i/2}}
\end{align}
where the Chern roots are $ch(TM)=\sum_i e^{x_i}$.
This $k$ instanton partiton function is just the arithmetic or Todd genus of the moduli space
except an irrelevant  coefficient
\begin{align}
Z_{\textrm{pure},\,k,\, 5D}
=\sum_n (-1)^n \dim_{\mathbb{C}} H^{0,n}(\mathcal{M}_k).
\end{align}

In order to take the twisted boundary condition into account,
we have to evaluate the following partition function by using the equivariant localization theorem
\begin{align}
Z_{\textrm{pure}, 5D}=\sum_{k=0}^\infty
q^k
\int_{\mathcal{M}_k}
\prod_{i}\frac{Rx_i}{1-e^{-Rx_i}}.
\end{align}
It is known that the $U(1)^{N_c-1}\times U(1)^2$,
which is the maximal torus of the isometry of the ADHM instanton moduli space,
enable us to apply the equivariant localization method to compute this partition function.
The Duistermaat-Heckman formula leads to
\begin{align}
Z_{\textrm{pure}, 5D}=\sum_{k=0}^\infty
q^k
\sum_{p:\textrm{fixed point}}
\prod_{i}\frac{1}{1-e^{-Rw_{i,p}}},
\end{align}
where $w_{i,p}$ are weights of the torus action at the fixed point $p$.
Fixed points of the torus action $U(1)^{N_c-1}\times U(1)^2$ are labelled by the Young diagrams $\vec{Y}
=(Y_1,\cdots,Y_{N_c})$
which satisfy
$Y_1+\cdots+ Y_{N_c}=k$.
For the fixed point $\vec{Y}$, the weights were computed by Nakajima \cite{NakajimaLecture:1999} as
\begin{align}
\label{NakajimaChern}
\sum_{i}e^{-Rw_{i,p{\scriptsize (\vec{Y})} }}
=\sum_{a,b=1}^{N_c}e^{-R(a_a-a_b)}
\Big(
\sum_{(i,j)\in Y_a}q^{-Y_{bj}^T+i}\,t^{-Y_{ai}+j-1}
+\sum_{(i,j)\in Y_b}q^{Y_{aj}^T-i+1}\,t^{Y_{bi}-j}
\Big),
\end{align}
where $q=e^{-R\epsilon_1}$ and $t=e^{R\epsilon_2}$.
There are thus $2N_c|\vec{Y}|=2kN_c$ fixed points 
which contribute to the instanton partition function via the localization theorem.
Notice that this number $2kN_c$ is precisely the dimensions of the ADHM moduli space of $k$-instantons.
By substituting these weights, we arrive at the partition function of the $\mathcal{N}=2$ pure super Yang-Mills theory
\cite{Nekrasov:2002qd,Flume:2002az,Bruzzo:2002xf,Nakajima:2003}:
\begin{align}
Z_{\textrm{pure}, 5D}=\sum_{k=0}^\infty
q^k
\sum_{|\vec{Y}|=k}
\prod_{a,b=1}^{N_c}\prod_{(i,j)\in Y_a}\frac{1}{1-Q_{ab}q^{-Y_{bj}^T+i}\,t^{-Y_{ai}+j-1}}
\prod_{(i,j)\in Y_b}\frac{1}{1-Q_{ab}q^{Y_{aj}^T-i+1}\,t^{Y_{bi}-j}},
\end{align}
where $Q_{ab}=e^{-R(a_a-a_b)}$.
The contribution of a vector multiplet is
\begin{align}
z_{\textrm{vect.}, 5D}(\vec{a},\vec{Y};q,t)=
\prod_{a,b=1}^{N_c}\prod_{(i,j)\in Y_a}\frac{1}{1-Q_{ab}q^{-Y_{bj}^T+i}\,t^{-Y_{ai}+j-1}}
\prod_{(i,j)\in Y_b}\frac{1}{1-Q_{ab}q^{Y_{aj}^T-i+1}\,t^{Y_{bi}-j}}.
\end{align}

\subsection*{Hyper multiplet}
In order to introduce matter hyper multiplets, 
we consider the matter bundle $E$ over the instanton moduli space $\mathcal{M}_k$
whose fiber is the Dirac zero modes  on the instanton background 
corresponding to the base point of $\mathcal{M}_k$.
Then the instanton partition function is the index for the twisted complex
\begin{align}
Z_{k ;\, 5D}&=\textrm{Ind}\,\mathcal{D}_E\nonumber\\
&=\int_{\mathcal{M}_k}
\hat{A}(T\mathcal{M}_k) ch(E).
\end{align}
The matter contribution therefore leads to such insertion 
of the equivariant Euler characteristic into the partition function.

Since the adjoint matter bundle is just the tangent bundle to $\mathcal{M}_k$,
the instanton partition function for $\mathcal{N}=2^*$ theory is
\begin{align}
Z_{\textrm{adj}, 5D}=\sum_{k=0}^\infty
q^k
\int_{\mathcal{M}_k}
\prod_{i}(1-e^{-Rm}e^{-Rx_i})\frac{Rx_i}{1-e^{-Rx_i}}
=\sum_{k=0}^\infty
q^k
\chi_{y}(\mathcal{M}_k),
\end{align}
where $y=e^{-Rm}$.
This is precisely the $\chi_{y}$ genus of the moduli space.
When $m=0$ the $k$-instanton partition function $\chi_{y}$ reduces to the Euler
characteristic $\sum (-1)^{n,m} \dim_{\mathbb{C}} H^{n,m}(\mathcal{M}_k)$.  
Note that this limit gives the partition function for $\mathcal{N}=4$ Yang-Mills theory:
\begin{align}
Z_{\mathcal{N}=4,\, 5D}=\sum_{k=0}^\infty
q^k
\chi(\mathcal{M}_k),
\end{align}
Let us compute the instanton partition function of the $\mathcal{N}=2^*$ theory with nonzero mass.
The Duistermaat-Heckman formula leads to
\begin{align}
Z_{\textrm{\,adj}, 5D}=\sum_{k=0}^\infty
\,q^k
\sum_{p\in\textrm{fixed points}}
\prod_{i}(1-e^{-Rm}e^{-Rw_{i,p}})\frac{1}{1-e^{-Rw_{i,p}}},
\end{align}
By using the Chern character (\ref{NakajimaChern}), 
we arrive at the instanton partition function of the $\mathcal{N}=2^*$ theory.
The contribution of the adjoint hypermultiplet is 
\begin{align}
z_{\textrm{adj.}, 5D}(\vec{a},m,\vec{Y};q,t)&=
\prod_{a,b=1}^{N_c}\prod_{(i,j)\in Y_a}\left({1-e^{-R(a_a-a_b-m)}q^{-Y_{bj}^T+i}\,t^{-Y_{ai}+j-1}}\right)\nonumber\\
&\qquad\qquad \times\prod_{(i,j)\in Y_b}\left({1-e^{-R(a_a-a_b-m)}q^{Y_{aj}^T-i+1}\,t^{Y_{bi}-j}}\right),
\end{align}

We can also discuss the contributions of matters in other representations using the localization method.
The results for the fundamental and bifundamental representations are
\begin{align}
&z_{\textrm{\,fund.}, 5D}(\vec{a},m,\vec{Y};q,t)=
\prod_{a=1}^{N_c}\prod_{(i,j)\in Y_a}\left({1-e^{-R(-a_a-m)}q^{i-1}\,t^{-j+1}}\right),\nonumber\\
&z_{\textrm{bifund.}, 5D}(\vec{a},\vec{Y};\vec{b},\vec{W};q,t)=
\prod_{a,b=1}^{N_c}\prod_{(i,j)\in Y_a}\left({1-e^{-R(a_a-b_b-m)}q^{-W_{bj}^T+i}\,t^{-Y_{ai}+j-1}}\right)\nonumber\\
&\qquad\qquad\qquad\qquad\qquad\qquad\qquad
 \times\prod_{(i,j)\in W_b}\left({1-e^{-R(a_a-b_b-m)}q^{Y_{aj}^T-i+1}\,t^{W_{bi}-j}}\right).
\end{align}
When one consider a gauge theory with matters in these representations,
one can construct the partition function by using these factors.

\subsection{Useful formulae}
Let us introduce the following combinatorical factor 
\begin{align}
N_{Y_a,Y_b}(Q;q,t):=
\prod_{s\in Y_a}\left({1-Qq^{a_{Y_a}(s)}\,t^{l_{Y_b}(t)+1}}\right)
\prod_{t\in Y_b}\left({1-Qq^{-a_{Y_b}(t)-1}\,t^{-l_{Y_a}(t)}}\right).
\end{align}
Note that this is a basic building block of the instanton partition functions.
The following identities are very important and  ubiquitous
in the study of the Nekrasov partition functions:
\begin{align}
N_{Y_a,Y_b}(Q;q,t)&=
\prod_{s\in Y_b}\left({1-Qq^{a_{Y_a}(s)}\,t^{l_{Y_b}(s)+1}}\right)
\prod_{t\in Y_a}\left({1-Qq^{-a_{Y_b}(t)1}\,t^{-l_{Y_a}(t)}}\right)\nonumber\\
&=\prod_{i,j=1}^\infty \frac{1-Qq^{-Y_{bi} +j-1}t^{-Y^T_{aj}+i} }
{1-Qq^{ j-1}t^{i}}\nonumber\\
&=\prod_{i,j=1}^\infty \frac{1-Qq^{Y_{ai} -j}t^{Y^T_{bj}-i+1} }
{1-Qq^{ -j}t^{-i+1}}.
\end{align}
Under the transposition of the Young diagrams, this satisfies
\begin{align}
N_{Y_a,Y_b}(Q^{-1};q,t)=
N_{Y_a^T,Y_b^T}(Q;t,q)\,
\frac{f_{Y_a}(q,t)}{f_{Y_b}(q,t)}
\left(Q\sqrt{\frac{q}{t}} \right)^{-|Y_a|-|Y_b|},
\end{align}
where we introduce the modified framing factor of \cite{Taki:2007dh} 
$f_{Y}(q,t)=(-1)^{|Y|}q^{||Y||^2/2}\,t^{-||Y^T||^2/2}$.
We then get the following property of the vector multiplet factor:
\begin{align}
\prod_{a,b=1}^{N_c}N_{Y_a,Y_b}(Q_{ab};t^{-1},q^{-1})=
\prod_{a,b=1}^{N_c}N_{Y_a^T,Y_b^T}(Q_{ab};q,t)=
\left( {\frac{t}{q}}\right)^{N_c|\vec{Y}|}\prod_{a,b=1}^{N_c}N_{Y_a,Y_b}(Q_{ab}^{-1};t,q)
\end{align}
Consult the extensive study of Awata and Kanno \cite{Awata:2005fa,Awata:2008ed} 
for proof of these identities and other combinatorical properties of the Nekrasov partition functions.

By using these formulae, we can derive the following expression for the instanton measures \cite{Taki:2007dh}:
\begin{align}
z_{\textrm{vect.}, 5D}(\vec{a},\vec{Y};q,t)&=(-1)^{N_c|\vec{Y}|}
\left( \frac{t}{q}\right)^{\frac{1}{2}N_c|\vec{Y}|}
\frac{t^{\frac{1}{2}\sum_a (N_c-2a+1)||Y_a^T||^2}}
{q^{\frac{1}{2}\sum_a (N_c-2a+1)||Y_a||^2}}
\prod_{1\leq a<b\leq N_c}{Q_{ab}}^{|Y_a|+|Y_b|}
\nonumber\\
&\quad\times
\prod_{a=1}^{N_c}q^{\frac{||Y_a||^2}{2}}\tilde{Z}_{Y_a}(t,q)
t^{\frac{||Y_a^T||^2}{2}}\tilde{Z}_{Y_a^T}(q,t)
\nonumber\\
&\quad\times \prod_{a<b}
\prod_{i,j=1}^\infty \frac{1-Q_{ab}q^{ -i}t^{-j+1}}{1-Q_{ab}q^{Y_{aj}^T-i}t^{Y_{bi}-j+1} }
\prod_{i,j=1}^\infty \frac{1-Q_{ab}q^{ -i+1}t^{-j}}{1-Q_{ab}q^{Y_{aj}^T -i+1}t^{Y_{bi}-j} }.
\end{align}
It is this identity of which we have to make good use 
in rewriting the refined topological string amplitudes into the Nekrasov partition functions.

The instanton measure of hypermultiplets also have infinite-product expressions:
\begin{align}
&z_{\textrm{adj.}, 5D}(\vec{a},m,\vec{Y};q,t)=
\prod_{a,b=1}^{N_c}
\prod_{i,j=1}^\infty \frac{1-Q_{ab}Q_m^{-1}q^{Y_{aj}^T-i+1}t^{Y_{bi}-j} }{1-Q_{ab}Q_m^{-1}q^{ -i+1}t^{-j}},
\nonumber\\
&z_{\textrm{\,fund.}, 5D}(\vec{a},m,\vec{Y};q,t)=
\prod_{a=1}^{N_c}
\prod_{i,j=1}^\infty \frac{1-e^{-R(-a_a-m)}q^{Y_{aj}^T-i}t^{-j+1} }{1-e^{-R(-a_a-m)}q^{ -i}t^{-j+1}},
\nonumber\\
&z_{\textrm{bifund.}, 5D}(\vec{a},\vec{Y};\vec{b},\vec{W};q,t)=
\prod_{a,b=1}^{N_c}
\prod_{i,j=1}^\infty \frac{1-Q_{ab}Q_m^{-1}q^{Y_{aj}^T-i+1}t^{W_{bi}-j} }{1-Q_{ab}Q_m^{-1}q^{ -i+1}t^{-j}}.
\end{align}
We utilize these formulae for rewriting topological string partition function
into Nekrasov instanton partition functions.

\subsection{Refined topological vertex}

Inspired by the connection between the topological vertex and the Nekrasov partition functions \cite{Eguchi:2003it},
some groups seek
an extended topological string theory corresponding to the general $\Omega$-backgrounds.
The refined topological vertex which was proposed by Iqbal, Kozcaz and Vafa in \cite{Iqbal:2007ii} was
based on the melting crystal dual picture of the A-model\footnote[2]{An another refined vertex formalism
had been proposed by Awata and Kanno \cite{Awata:2005fa}.
They employed the Macdonald functions for refinement.
Since their results agree with the closed string amplitudes of \cite{Iqbal:2007ii},
it is believed that these two formulations are equivalent mathematically \cite{Awata:2008ed}.}.
The precise form of the vertex function is
\begin{align}
&C_{Y_1 Y_2 Y_3 } (t,q)\nonumber\\
&\rule{0pt}{5ex}
={\left( {\frac{q}{t}} \right)}^{\frac{{\left\| Y_2  \right\|^2  + \left\| Y_3  \right\|^2 }}{2}} t^{\frac{{{\kappa}_{Y_2}  }}{2}} 
P_{{Y_3}^T } (t^{ - \rho } ;q,t)
\sum_Y  {{\left( {\frac{q}{t}} \right)}^{\frac{{\left| Y  \right| + \left| Y_1  \right| - \left| Y_2  \right|}}{2}} 
S_{{Y_1}^T /Y } (t^{ - \rho } q^{ - Y_3 } ) 
S_{Y_2 /Y } (t^{ - {Y_3}^T } q^{ - \rho } )},
\end{align}
where $P_{{Y}^T } (\mathop t\nolimits^{ - \rho } ;q,t)$ is the following specialization
of the Macdonald function \cite{macdonald,Awata:2005fa}
\begin{align}
P_{Y^T } (t^{ - \rho } ;q,t)=t^{\frac{1}{2}||Y||^2}\tilde{Z}_Y(t,q),
\quad
\tilde{Z}_{Y}(t,q)=\prod_{(i,j)\in Y}(1-t^{{Y_j}^T-i+1} q^{Y_i-j})^{-1}.
\end{align} 
The arguments $x=t^{ - \rho } q^{ - Y}$ denote $x_i=t^{ i-1/2 } q^{ - Y_i}$ for $i=1,2,\cdots$.



\begin{thebibliography}{99}


\bibitem{Seiberg:1994rs}
  N.~Seiberg and E.~Witten,
  ``Monopole Condensation, And Confinement In N=2 Supersymmetric Yang-Mills
  Theory,''
  Nucl.\ Phys.\  B {\bf 426}, 19 (1994)
  [Erratum-ibid.\  B {\bf 430}, 485 (1994)]
  [arXiv:hep-th/9407087].



\bibitem{Seiberg:1994aj}
  N.~Seiberg and E.~Witten,
  ``Monopoles, duality and chiral symmetry breaking in N=2 supersymmetric
  QCD,''
  Nucl.\ Phys.\  B {\bf 431}, 484 (1994)
  [arXiv:hep-th/9408099].


\bibitem{Nekrasov:2002qd}
  N.~A.~Nekrasov,
  ``Seiberg-Witten Prepotential From Instanton Counting,''
  Adv.\ Theor.\ Math.\ Phys.\  {\bf 7}, 831 (2004)
  [arXiv:hep-th/0206161].
  
  
  

\bibitem{Alday:2009aq}
  L.~F.~Alday, D.~Gaiotto and Y.~Tachikawa,
  ``Liouville Correlation Functions from Four-dimensional Gauge Theories,''
  arXiv:0906.3219 [hep-th].



\bibitem{Gaiotto:2009we}
  D.~Gaiotto,
  ``N=2 dualities,''
  arXiv:0904.2715 [hep-th].


\bibitem{Drukker:2009tz}
  N.~Drukker, D.~R.~Morrison and T.~Okuda,
  ``Loop operators and S-duality from curves on Riemann surfaces,''
  JHEP {\bf 0909}, 031 (2009)
  [arXiv:0907.2593 [hep-th]].


\bibitem{Alday:2009fs}
  L.~F.~Alday, D.~Gaiotto, S.~Gukov, Y.~Tachikawa and H.~Verlinde,
  ``Loop and surface operators in N=2 gauge theory and Liouville modular
  geometry,''
  JHEP {\bf 1001}, 113 (2010)
  [arXiv:0909.0945 [hep-th]].


\bibitem{Drukker:2009id}
  N.~Drukker, J.~Gomis, T.~Okuda and J.~Teschner,
  ``Gauge Theory Loop Operators and Liouville Theory,''
  JHEP {\bf 1002}, 057 (2010)
  [arXiv:0909.1105 [hep-th]].



\bibitem{Ooguri:1999bv}
  H.~Ooguri and C.~Vafa,
  ``Knot invariants and topological strings,''
  Nucl.\ Phys.\  B {\bf 577}, 419 (2000)
  [arXiv:hep-th/9912123].
  

  \bibitem{Katz:1996fh}
  S.~H.~Katz, A.~Klemm and C.~Vafa,
  ``Geometric engineering of quantum field theories,''
  Nucl.\ Phys.\  B {\bf 497}, 173 (1997)
  [arXiv:hep-th/9609239].

\bibitem{Aganagic:2003qj}
  M.~Aganagic, R.~Dijkgraaf, A.~Klemm, M.~Marino and C.~Vafa,
  ``Topological strings and integrable hierarchies,''
  Commun.\ Math.\ Phys.\  {\bf 261}, 451 (2006)
  [arXiv:hep-th/0312085].




\bibitem{Dimofte:2010tz}
  T.~Dimofte, S.~Gukov and L.~Hollands,
  ``Vortex Counting and Lagrangian 3-manifolds,''
  arXiv:1006.0977 [hep-th].












  
  
  
  
 \bibitem{Wyllard:2009hg}
  N.~Wyllard,
  ``$A_{N-1}$ conformal Toda field theory correlation functions from conformal
  N=2 SU(N) quiver gauge theories,''
  JHEP {\bf 0911}, 002 (2009)
  [arXiv:0907.2189 [hep-th]].
   
  \bibitem{Schiappa:2009cc}
  R.~Schiappa and N.~Wyllard,
  ``An $A_r$ threesome: Matrix models, 2d CFTs and 4d N=2 gauge theories,''
  arXiv:0911.5337 [hep-th].


\bibitem{Kozcaz:2010af}
  C.~Kozcaz, S.~Pasquetti and N.~Wyllard,
  ``A \& B model approaches to surface operators and Toda theories,''
  arXiv:1004.2025 [hep-th].
  
  
  
  
  \bibitem{Iqbal:2002we}
  A.~Iqbal,
  ``All genus topological string amplitudes and 5-brane webs as Feynman
  diagrams,''
  arXiv:hep-th/0207114.


\bibitem{Iqbal:2003ix}
  A.~Iqbal and A.~K.~Kashani-Poor,
  ``Instanton counting and Chern-Simons theory,''
  Adv.\ Theor.\ Math.\ Phys.\  {\bf 7}, 457 (2004)
  [arXiv:hep-th/0212279].


\bibitem{Iqbal:2003zz}
  A.~Iqbal and A.~K.~Kashani-Poor,
  ``SU(N) geometries and topological string amplitudes,''
  Adv.\ Theor.\ Math.\ Phys.\  {\bf 10}, 1 (2006)
  [arXiv:hep-th/0306032].



\bibitem{Eguchi:2003sj}
  T.~Eguchi and H.~Kanno,
  ``Topological strings and Nekrasov's formulas,''
  JHEP {\bf 0312}, 006 (2003)
  [arXiv:hep-th/0310235].
  
\bibitem{Hollowood:2003cv}
  T.~J.~Hollowood, A.~Iqbal and C.~Vafa,
  ``Matrix Models, Geometric Engineering and Elliptic Genera,''
  JHEP {\bf 0803}, 069 (2008)
  [arXiv:hep-th/0310272].  
  
\bibitem{Eguchi:2003it}
  T.~Eguchi and H.~Kanno,
  ``Geometric transitions, Chern-Simons gauge theory and Veneziano type
  amplitudes,''
  Phys.\ Lett.\  B {\bf 585}, 163 (2004)
  [arXiv:hep-th/0312234].






\bibitem{Aganagic:2003db}
  M.~Aganagic, A.~Klemm, M.~Marino and C.~Vafa,
  ``The topological vertex,''
  Commun.\ Math.\ Phys.\  {\bf 254}, 425 (2005)
  [arXiv:hep-th/0305132].
  
  
  \bibitem{KashaniPoor:2006nc}
  A.~K.~Kashani-Poor,
  ``The Wave Function Behavior of the Open Topological String Partition
  Function on the Conifold,''
  JHEP {\bf 0704}, 004 (2007)
  [arXiv:hep-th/0606112].
  
  \bibitem{KashaniPoor:2008xg}
  A.~K.~Kashani-Poor,
  ``Phase space polarization and the topological string: a case study,''
  Mod.\ Phys.\ Lett.\  A {\bf 23}, 3199 (2008)
  [arXiv:0812.0687 [hep-th]].
  
  
\bibitem{Witten:1997sc}
  E.~Witten,
  ``Solutions of four-dimensional field theories via M-theory,''
  Nucl.\ Phys.\  B {\bf 500}, 3 (1997)
  [arXiv:hep-th/9703166].  
  

\bibitem{Iqbal:2007ii}
  A.~Iqbal, C.~Kozcaz and C.~Vafa,
  ``The refined topological vertex,''
  JHEP {\bf 0910}, 069 (2009)
  [arXiv:hep-th/0701156].



\bibitem{Awata:2009ur}
  H.~Awata and Y.~Yamada,
  ``Five-dimensional AGT Conjecture and the Deformed Virasoro Algebra,''
  JHEP {\bf 1001}, 125 (2010)
  [arXiv:0910.4431 [hep-th]].



\bibitem{Gukov:2007tf}
  S.~Gukov, A.~Iqbal, C.~Kozcaz and C.~Vafa,
  ``Link homologies and the refined topological vertex,''
  arXiv:0705.1368 [hep-th].


\bibitem{Nekrasov:2009rc}
  N.~A.~Nekrasov and S.~L.~Shatashvili,
  ``Quantization of Integrable Systems and Four Dimensional Gauge Theories,''
  arXiv:0908.4052 [hep-th].


  
  
  
  
  

\bibitem{Mironov:2009uv}
  A.~Mironov and A.~Morozov,
  ``Nekrasov Functions and Exact Bohr-Sommerfeld Integrals,''
  JHEP {\bf 1004}, 040 (2010)
  [arXiv:0910.5670 [hep-th]].


\bibitem{Mironov:2009dv}
  A.~Mironov and A.~Morozov,
  ``Nekrasov Functions from Exact BS Periods: the Case of SU(N),''
  J.\ Phys.\ A  {\bf 43}, 195401 (2010)
  [arXiv:0911.2396 [hep-th]].


\bibitem{Popolitov:2010bz}
  A.~Popolitov,
  ``On relation between Nekrasov functions and BS periods in pure SU(N) case,''
  arXiv:1001.1407 [hep-th].
  
\bibitem{Maruyoshi:2010iu}
  K.~Maruyoshi and M.~Taki,
  ``Deformed Prepotential, Quantum Integrable System and Liouville Field
  Theory,''
  arXiv:1006.4505 [hep-th].
  

 

  
  
  
  
  




  
  
  
  
  
  
  
  
  
  
  
  
  

\bibitem{Ruback:1986ag}
  P.~J.~Ruback,
  ``THE MOTION OF KALUZA-KLEIN MONOPOLES,''
  Commun.\ Math.\ Phys.\  {\bf 107}, 93 (1986).

\bibitem{Ooguri:1995wj}
  H.~Ooguri and C.~Vafa,
  ``Two-Dimensional Black Hole and Singularities of CY Manifolds,''
  Nucl.\ Phys.\  B {\bf 463}, 55 (1996)
  [arXiv:hep-th/9511164].


\bibitem{Dijkgraaf:2007sw}
  R.~Dijkgraaf, L.~Hollands, P.~Sulkowski and C.~Vafa,
  ``Supersymmetric Gauge Theories, Intersecting Branes and Free Fermions,''
  JHEP {\bf 0802}, 106 (2008)
  [arXiv:0709.4446 [hep-th]].
  
\bibitem{Hori:2000kt}
  K.~Hori and C.~Vafa,
  ``Mirror symmetry,''
  arXiv:hep-th/0002222.


\bibitem{Aganagic:2000gs}
  M.~Aganagic and C.~Vafa,
  ``Mirror symmetry, D-branes and counting holomorphic discs,''
  arXiv:hep-th/0012041.

\bibitem{Aganagic:2001nx}
  M.~Aganagic, A.~Klemm and C.~Vafa,
  ``Disk instantons, mirror symmetry and the duality web,''
  Z.\ Naturforsch.\  A {\bf 57}, 1 (2002)
  [arXiv:hep-th/0105045]
  
  
\bibitem{Gopakumar:1998ii}
  R.~Gopakumar and C.~Vafa,
  ``M-theory and topological strings. I,''
  arXiv:hep-th/9809187.

\bibitem{Gopakumar:1998ki}
  R.~Gopakumar and C.~Vafa,
  ``On the gauge theory/geometry correspondence,''
  Adv.\ Theor.\ Math.\ Phys.\  {\bf 3}, 1415 (1999)
  [arXiv:hep-th/9811131].

\bibitem{Gopakumar:1998jq}
  R.~Gopakumar and C.~Vafa,
  ``M-theory and topological strings. II,''
  arXiv:hep-th/9812127.
  
  
  
\bibitem{Dijkgraaf:2009pc}
  R.~Dijkgraaf and C.~Vafa,
  ``Toda Theories, Matrix Models, Topological Strings, and N=2 Gauge Systems,''
  arXiv:0909.2453 [hep-th].
  
\bibitem{Mironov:2010zs}
  A.~Mironov, A.~Morozov and S.~Shakirov,
  ``Conformal blocks as Dotsenko-Fateev Integral Discriminants,''
  Int.\ J.\ Mod.\ Phys.\  A {\bf 25}, 3173 (2010)
  [arXiv:1001.0563 [hep-th]].

\bibitem{Itoyama:2010ki}
  H.~Itoyama and T.~Oota,
  ``Method of Generating q-Expansion Coefficients for Conformal Block and N=2
  Nekrasov Function by beta-Deformed Matrix Model,''
  Nucl.\ Phys.\  B {\bf 838}, 298 (2010)
  [arXiv:1003.2929 [hep-th]].

\bibitem{Itoyama:2009sc}
  H.~Itoyama, K.~Maruyoshi and T.~Oota,
  ``Notes on the Quiver Matrix Model and 2d-4d Conformal Connection,''
  arXiv:0911.4244 [hep-th].

\bibitem{Eguchi:2009gf}
  T.~Eguchi and K.~Maruyoshi,
  ``Penner Type Matrix Model and Seiberg-Witten Theory,''
  JHEP {\bf 1002}, 022 (2010)
  [arXiv:0911.4797 [hep-th]].



\bibitem{Taki:2007dh}
  M.~Taki,
  ``Refined Topological Vertex and Instanton Counting,''
  JHEP {\bf 0803}, 048 (2008)
  [arXiv:0710.1776 [hep-th]].



\bibitem{Iqbal:2004ne}
  A.~Iqbal and A.~K.~Kashani-Poor,
  ``The vertex on a strip,''
  Adv.\ Theor.\ Math.\ Phys.\  {\bf 10}, 317 (2006)
  [arXiv:hep-th/0410174].



\bibitem{Gomis:2006mv}
  J.~Gomis and T.~Okuda,
  ``Wilson loops, geometric transitions and bubbling Calabi-Yau's,''
  JHEP {\bf 0702}, 083 (2007)
  [arXiv:hep-th/0612190].
  
\bibitem{Gomis:2007kz}
  J.~Gomis and T.~Okuda,
  ``D-branes as a Bubbling Calabi-Yau,''
  JHEP {\bf 0707}, 005 (2007)
  [arXiv:0704.3080 [hep-th]].


  
  

\bibitem{Marshakov:2009gs}
  A.~Marshakov, A.~Mironov and A.~Morozov,
  ``On Combinatorial Expansions of Conformal Blocks,''
  arXiv:0907.3946 [hep-th].

\bibitem{Mironov:2009dr}
  A.~Mironov, S.~Mironov, A.~Morozov and A.~Morozov,
  ``CFT exercises for the needs of AGT,''
  arXiv:0908.2064 [hep-th].

  
\bibitem{Alba:2009fp}
  V.~Alba and A.~Morozov,
  ``Non-conformal limit of AGT relation from the 1-point torus conformal
  block,''
  arXiv:0911.0363 [hep-th].
  
  
  \bibitem{Alba:2009ya}
  V.~Alba and A.~Morozov,
  ``Check of AGT Relation for Conformal Blocks on Sphere,''
  arXiv:0912.2535 [hep-th].
  




  
  


\bibitem{Gaiotto:2009ma}
  D.~Gaiotto,
  ``Asymptotically free N=2 theories and irregular conformal blocks,''
  arXiv:0908.0307 [hep-th].
  
\bibitem{Marshakov:2009gn}
  A.~Marshakov, A.~Mironov and A.~Morozov,
  ``On non-conformal limit of the AGT relations,''
  Phys.\ Lett.\  B {\bf 682}, 125 (2009)
  [arXiv:0909.2052 [hep-th]].  


\bibitem{Gaiotto:2009fs}
  D.~Gaiotto,
  ``Surface Operators in N=2 4d Gauge Theories,''
  arXiv:0911.1316 [hep-th].


\bibitem{Alday:2010vg}
  L.~F.~Alday and Y.~Tachikawa,
  ``Affine SL(2) conformal blocks from 4d gauge theories,''
  arXiv:1005.4469 [hep-th].
  
\bibitem{Mironov:2009by}
  A.~Mironov and A.~Morozov,
  ``On AGT relation in the case of U(3),''
  Nucl.\ Phys.\  B {\bf 825}, 1 (2010)
  [arXiv:0908.2569 [hep-th]].  
  
\bibitem{Taki:2009zd}
  M.~Taki,
  ``On AGT Conjecture for Pure Super Yang-Mills and W-algebra,''
  arXiv:0912.4789 [hep-th].
  

\bibitem{Teschner}
  J.~Teschner,
  ``Quantization of the Hitchin moduli spaces, Liouville theory, and the
  geometric Langlands correspondence,''
  arXiv:1005.2846 [hep-th].
  
  \bibitem{Bonelli:2009zp}
  G.~Bonelli and A.~Tanzini,
  ``Hitchin systems, N=2 gauge theories and W-gravity,''
  arXiv:0909.4031 [hep-th].
 


 \bibitem{Flume:2002az}
  R.~Flume and R.~Poghossian,
  ``An algorithm for the microscopic evaluation of the coefficients of the
  Seiberg-Witten prepotential,''
  Int.\ J.\ Mod.\ Phys.\  A {\bf 18}, 2541 (2003)
  [arXiv:hep-th/0208176].
  
 \bibitem{Bruzzo:2002xf}
  U.~Bruzzo, F.~Fucito, J.~F.~Morales and A.~Tanzini,
  ``Multi-instanton calculus and equivariant cohomology,''
  JHEP {\bf 0305}, 054 (2003)
  [arXiv:hep-th/0211108].
  
\bibitem{Nakajima:2003}
  H.~Nakajima and K.~Yoshioka,
  ``Instanton counting on blowup. I. 4-dimensional pure gauge theory,''
  Invent.\ Math {\bf 162}, no. 2, 313 (2005)
  [arXiv:math.A.G/0306198].
  
\bibitem{NakajimaLecture:1999}
H.~Nakajima, 
“Lectures on Hilbert Schemes of Points on Surfaces”, 
American Mathematical Society,University Lectures Series v.18 (1999)



\bibitem{Awata:2005fa}
  H.~Awata and H.~Kanno,
  ``Instanton counting, Macdonald functions and the moduli space of
  D-branes,''
  JHEP {\bf 0505}, 039 (2005)
  [arXiv:hep-th/0502061].
 


\bibitem{Awata:2008ed}
  H.~Awata and H.~Kanno,
  ``Refined BPS state counting from Nekrasov's formula and Macdonald
  functions,''
  Int.\ J.\ Mod.\ Phys.\  A {\bf 24}, 2253 (2009)
  [arXiv:0805.0191 [hep-th]].


\bibitem{macdonald}
  I. G. Macdonald, ``Symmetric functions and Hall polynomials,''
(second edition, 1995), Oxford Mathematical Monographs,
 Oxford Science Publications.

\end{thebibliography}
\end{document}